\documentclass[preprint,12pt,review]{elsarticle}
\usepackage[utf8]{inputenc}
\usepackage{graphicx}
\usepackage{amssymb}
\usepackage{soul}
\usepackage{appendix}
\usepackage{amsthm}
\usepackage{setspace}
\usepackage{amsmath}
\makeatletter

\renewcommand\tagform@[1]{\maketag@@@{\ignorespaces#1\unskip\@@italiccorr}}
\renewcommand\theequation{(\oldtheequation)}
\makeatother
\usepackage{amssymb}
\usepackage{textcomp}
\usepackage{gensymb}
\usepackage{xspace}
\usepackage{longtable}
\usepackage{booktabs}
\usepackage[normalem]{ulem}
\usepackage[version=3]{mhchem} 
\usepackage[usenames]{color}
\usepackage[T1]{fontenc}
\usepackage{xr-hyper} 
\usepackage[pdfpagelabels,bookmarks]{hyperref}
\usepackage{upgreek}
\usepackage{siunitx}
\usepackage{caption}
\usepackage{pifont}
\usepackage[english]{babel}
\usepackage{geometry} 
\biboptions{sort&compress,comma}

\addto\extrasenglish{

}
\renewcommand{\eqref}[1]{Eq.~\ref{#1}}

\newcommand{\figsref}[2]{Figures~\ref{#1} and \ref{#2}}
\newcommand{\eqsref}[2]{Eqs.~\ref{#1}~\&~\ref{#2}}
\newcommand{\eqssref}[3]{Eqs.~\ref{#1},~\ref{#2}~\&~\ref{#3}}

\newcommand\unappendix{\par
  \setcounter{section}{0}%
  \setcounter{subsection}{0}%
  \gdef\thesection{\@arabic\c@section}}
  \renewcommand{\appendixautorefname}{Section}
\makeatother

\usepackage{newfloat}
\DeclareFloatingEnvironment[
  fileext   = los,                      
  listname  = {List of Files},          
  name      = File,                     
  placement = htp                       
]{file}

\begin{document}

\begin{frontmatter}

\title{Understanding Electrolyte Filling of Lithium-Ion Battery Electrodes on the Pore Scale Using the Lattice Boltzmann Method}

 \author[DLR,HIU]{Martin P. Lautenschlaeger}\corref{1}
 \cortext[1]{Corresponding author:}
 \ead{Martin.Lautenschlaeger@dlr.de}
 \author[UUlmS]{Benedikt Prifling}
 \author[DLR,HIU]{Benjamin Kellers}
 \author[DLR,HIU]{Julius Weinmiller}
 \author[DLR,HIU]{Timo Danner}
 \author[UUlmS]{Volker Schmidt}
 \author[DLR,HIU,UUlmE]{Arnulf Latz}
 
 \address[DLR]{German Aerospace Center (DLR), Institute of Engineering Thermodynamics, 70569 Stuttgart, Germany}
 \address[HIU]{Helmholtz Institute Ulm for Electrochemical Energy Storage (HIU), 89081 Ulm, Germany}
 \address[UUlmS]{Ulm University (UUlm), Institute of Stochastics, 89081 Ulm, Germany}
 \address[UUlmE]{Ulm University (UUlm), Institute of Electrochemistry, 89081 Ulm, Germany}

\begin{abstract}
Electrolyte filling is a time-critical step during battery manufacturing that also affects the battery performance. The underlying physical phenomena during filling mainly occur on the pore scale and are hard to study experimentally. In this paper, a computational approach, i.e.\ the lattice Boltzmann method, is used to study the filling process and corresponding pore-scale phenomena in 3D lithium-ion battery cathodes. The electrolyte flow through the nanoporous binder is simulated using a homogenization approach. Besides the process time, the influence of structural and physico-chemical properties is investigated. Those are the particle size, the binder distribution, and the volume fraction and wetting behavior of active material and binder. Optimized filling conditions are discussed by capillary pressure-saturation relationships. It is shown how the aforementioned influencing factors affect the electrolyte saturation. Moreover, the amount of the entrapped residual gas phase and the corresponding size distribution of the gas agglomerates are analyzed in detail. Both factors are shown to have a strong impact on mechanisms that can adversely affect the battery performance. The results obtained here indicate how the filling process, the final electrolyte saturation, and potentially also the battery performance can be optimized by adapting process parameters and the electrode and electrolyte design.

\end{abstract}


\journal{arXiv}

\end{frontmatter}



\section{Introduction}

Lithium-ion batteries are the major power source for battery electric vehicles. Its cell production is predicted to increase exponentially in the upcoming years. Therefore, the optimization of the battery production is key to reduce costs and the environmental impact of next-generation battery cells. Improving the battery manufacturing process requires the optimization of each process step. One of the process steps that has recently gained attention in this context, is the filling of cells with liquid electrolyte, where the electrolyte is first dosed into the cell and subsequently stored to achieve a uniform electrolyte distribution. Sometimes this procedure is even repeated to decrease entrapment of residual gas \cite{Wood2015,Habedank2019}. Thus, the filling process is time-consuming and cost-intensive. It can take up to several days \cite{Wood2015,Knoche2016,Weydanz2018}. Additionally, the filling also affects the battery performance and lifetime \cite{Knoche2016,Knoche2016CIRP,Weydanz2018,Schilling2020}. It is known that poorly wetted pores in electrodes cause the development of inhomogeneous solid electrolyte interphases (SEI) \cite{Lanz2001}. Moreover, they can lead to electrolyte decomposition during cycling \cite{Imhof1998}, dendrite formation \cite{Wood2015,Knoche2016CIRP,Schilling2019}, and non-uniform current densities \cite{Mueller2018,Weydanz2018}. An incomplete wetting can also have a large effect on the battery performance by increasing internal ionic resistances remarkably, which has recently been investigated for separators \cite{Sauter2020}. 

There are different strategies to prevent the aforementioned limitations and to increase the wettability and the final degree of electrode saturation. Eihter the filling process is sped up by cell evacuation and applying pressure gradients \cite{Wu2004,Wood2015,Knoche2016,Knoche2016CIRP,Weydanz2018,Habedank2019,Guenter2022} or the physico-chemical properties of electrolyte and electrodes are tuned to improve the filling process \cite{Sauter2020}. Two main approaches are considered in the literature. Electrolyte properties, i.e.\ surface tension and viscosity, are adjusted by electrolyte additives \cite{Wu2004,Zhang2006,Guenter2022} or the electrode wettability is improved using coatings or surfactants \cite{Schilling2020}. Moreover, also structural properties of electrodes and separator are known to have a significant influence on the filling process \cite{Wood2015,Habedank2019,Schilling2019,Schilling2020,Davoodabadi2020,Sauter2020}. The compaction of the electrode by calendering, e.g., increases the filling duration and the amount of residual gas \cite{Wood2015,Habedank2019,Schilling2020,Davoodabadi2020}.

Recently, different experimental studies have investigated the filling process using in-situ methods. Amongst those were electrochemical impedance spectroscopy \cite{Guenter2018,Guenter2022}, neutron radiography \cite{Knoche2016,Weydanz2018,Habedank2019,Guenter2022}, X-ray measurements \cite{Schilling2019}, focused ion beam combined with scanning electron microscopy \cite{Davoodabadi2020}, thermography \cite{Schilling2020}, and wetting balance tests \cite{Schilling2020}. However, most of those methods are complex and time-consuming. They all suffer from low spatial or temporal resolution, imprecise localization of the wetting front, or cannot resolve the interdependency of the different influencing factors. Thus, a comprehensive understanding especially of pore-scale phenomena during the filling process is missing \cite{Davoodabadi2020}. There is still no common agreement on how to optimize this process, especially for the multitude of electrochemical systems and cell types available on the market \cite{Knoche2016CIRP}.

A method that is capable of giving a detailed insight into the wetting phenomena and the interdependency of the influencing factors are direct numerical simulations in general, and the lattice Boltzmann method (LBM) in particular. LBM has proven to be a reliable tool for the simulation of transport processes and fluid flow \cite{Chen1998,Krueger2016}. In contrast to conventional fluid dynamics, it gives access to multi-scale and multi-physics issues even within complex geometries, e.g., in porous media \cite{Chen2014,Liu2016}. 

The multi-component Shan-Chen pseudopotential method (MCSC) has regularly been utilized to simulate multi-phase flows with LBM \cite{Shan1993,Chen2014,Liu2016}. Similar to molecular dynamics simulations, where molecular interactions are modeled to study, e.g., wetting phenomena \cite{Diewald2020,Lautenschlaeger2020,Becker2014} or transport processes \cite{Lautenschlaeger2019,Lautenschlaeger2019a}, it uses fluid-fluid and solid-fluid interaction forces to model interfacial tension and adhesion forces, respectively \cite{Liu2016}. 

So far, LBM has been successfully applied to investigate water transport and hysteresis effects in catalyst or gas diffusion layers of polymer electrolyte membrane fuel cells \cite{Jeon2015,Satjaritanun2017,Sakaida2017,Niu2018,Jeon2020,Zhu2021,Grunewald2021}. However, it has rarely been applied in the context of battery simulations \cite{Jiang2016,Danner2016,Jiang2017,Jiang2018}. Only a few studies have been conducted in which LBM was applied to study electrolyte filling processes \cite{Lee2013,Lee2014,Jeon2019,Shodiev2021}. Jeon and co-workers \cite{Lee2013,Lee2014,Jeon2019} as well as Mohammadian and Zhang \cite{Mohammadian2018} studied the effect of structural properties and wettability on the filling duration. However, the underlying microstructures of the electrodes were rather simplified. Moreover, only 2D simulations were conducted, although this reduces the number of flow paths significantly and thereby strongly affects the saturation behavior, pore blocking, gas entrapment, and the simulation accuracy \cite{Danner2016,Jeon2019}. Electrolyte filling of realistic 3D lithium-ion battery electrodes using LBM was investigated only recently by Shodiev~\textit{et al.}\ \cite{Shodiev2021,Shodiev2021ML}. Their studies focused on the correlation between the structural properties of electrodes and the filling duration, from which the data were also used to train a machine learning algorithm. However, the wetting properties of active material and binder were assumed to be equal and the binder was fully solid and impermeable without considering its nanoporosity.

The present paper extends the findings of the aforementioned studies \cite{Shodiev2021,Shodiev2021ML}. In particular, the electrolyte filling process of realistic virtual 3D lithium-ion battery electrode structures is studied using LBM. The simulation setups and boundary conditions that are used mimic experimental setups. In addition, electrode structures and LBM model parameters are chosen such that they represent authentic materials typically used for lithium-ion batteries. Furthermore, motivated by the work of Pereira \cite{Pereira2016,Pereira2017,Pereira2019}, MCSC is combined with a homogenization approach that is based on the grayscale (GS) or partial bounce-back (PBB) method \cite{Walsh2009}. This allows to simultaneously study the electrolyte flow in the mesoscopic pores confined by active material particles as well as in the nanoscopic pores of the binder without structurally resolving the latter. Note that our model is applied to simulate electrolyte wetting in lithium-ion battery cathodes, but is not limited to this particular application. Other research fields can benefit from this development, e.g., flow phenomena in redox-flow batteries and fuel cells. Using our model, the process time as well as the influence of a wide range of relevant structural and physico-chemical properties of lithium-ion battery cathodes are studied. More precisely, the influence of the particle size distribution $R_{\text{PS}}$, the volume fraction $\phi_{\text{A}}$, and the wettability $\theta_{\text{A}}$ of the active material on the filling process is investigated. In addition, a permeable binder is virtually added to some of the electrodes for which the inner volume fraction $\phi'_{\text{B}}$ and the wettability $\theta_{\text{B}}$ are varied. 

This study aims to increase the understanding of the electrolyte filling process on the pore scale. It gives insight into the sensitivity of the aforementioned parameters on pressure-saturation behavior during filling and electrolyte saturation. For each electrode customized pressure profiles are determined that ensure a steady and uniform filling process. Finally, the amount and the size distribution of entrapped residual gas agglomerates are analyzed in detail. It is shown how the residual gas phase can adversely affect the battery performance. Moreover, permeabilities are determined to estimate the efforts for displacing gas agglomerates from the electrodes in a subsequent production step. All in all, the results presented here are helpful to optimize electrode and electrolyte design as well as the filling process. The findings are also applicable to optimize the filling of anodes, separators or other battery types.

The present paper is organized as follows. In \autoref{sec:LBM}, the LBM and the combination of MCSC and GS are described. The simulation setup is given in \autoref{sec:Setup}, where also the electrode structure generation and the analysis are described. \autoref{sec:StudyOverview} gives an overview of the study including the influencing factors. The results are presented in \autoref{sec:Results}. Finally, conclusions are drawn in \autoref{sec:Conclusion}.

\section{Methods}  \label{sec:LBM}

\subsection{Lattice Boltzmann Method}  \label{subsec:LBM}
Details regarding the background, derivation, and implementation of LBM are described in the literature \cite{Krueger2016}. A brief overview of the fundamentals of LBM, MCSC \cite{Shan1993}, and GS \cite{Walsh2009} is given in the Supporting Information (cf.\ \autoref{SI-sec:LBM}). In the following, the combination of MCSC with GS is described. It follows the approach developed by Pereira \cite{Pereira2016,Pereira2017,Pereira2019} which is adjusted here. 

The MCSC is used to model multi-phase flows. Interactions between phases and solid-fluid interactions with a solid wall are typically described using a pseudopotential. The GS is used to study physical situations in which the resolution of the numerical lattice is coarser than the smallest relevant physical length scale \cite{Schaap2007,He2019}, e.g.\ flow in pores which diameters differ by orders of magnitude. In the present work, MCSC and GS are combined to study multi-phase flows of liquid electrolyte and gas phase in electrodes consisting of mesopores confined by the active material and nanopores within the binder.

\subsection{Combining MCSC with GS} \label{subsec:Model} 
Similar to MCSC, in the model used here, each fluid component $\sigma$ is represented by a distinct distribution function $\boldsymbol{f}^{\sigma}\left(\mathbf{x}, t\right)$, where $\mathbf{x}\in\mathbb{R}^3$ and $t\geq0$ denote the position of the lattice cell and the time, respectively. It is discretized in velocity space on a regular cubic 3D lattice. Each lattice cell is linked to its 18 nearest neighbors, resulting in the so-called D3Q19 velocity set (cf.\ \autoref{eq:VelocitySet}). The links correspond to the directions $i$ along which the discrete distribution functions $f_i$ are streamed.

The temporal evolution of $\boldsymbol{f}$ is described by the lattice Boltzmann equation (cf.\ \autoref{SI-eq:LBM}). For the combined method, it reads
\begin{align}
    \label{eq:LGS}
    \begin{split}
    f^{\sigma}_i\left(\mathbf{x}+\mathbf{c}_i \Delta t, t + \Delta t \right) =&\quad\  (1-n_s^{\sigma}(\mathbf{x}))f^{\sigma}_i\left(\mathbf{x}, t\right) \\\
    &- (1-n_s^{\sigma}(\mathbf{x}))\frac{\Delta t}{\tilde{\tau}^{\sigma}} \left(f^{\sigma}_i\left(\mathbf{x}, t\right) - f^{\mathrm{eq},\sigma}_i\left(\mathbf{x}, t\right)\right) \\\
    &+ n_s^{\sigma}(\mathbf{x}) f^{\sigma}_{\bar{i}}\left(\mathbf{x}, t\right).
    \end{split}
\end{align}
The second line of \autoref{eq:LGS} describes the relaxation of $\boldsymbol{f}$ towards the Maxwell-Boltzmann equilibrium distribution function $\boldsymbol{f}^\mathrm{eq}$ (cf. \autoref{eq:Maxwell}). The characteristic relaxation time is $\tilde{\tau}$ and related to the kinematic viscosity via $\nu=c_{s}^{2}(\tilde{\tau}-1 / 2)$.  The parameter $\Delta t$ is the time step. The third line of \autoref{eq:LGS} is the bounce-back scheme (cf.\ \cite{Krueger2016}) representing no-slip conditions at solids \cite{Chen1998,Liu2016}. It corresponds to the state prior to the collision. The parameter $\bar{i}$ denotes the direction opposite to $i$ with the exception $i=0=\bar{i}$.

Furthermore, in \autoref{eq:LGS}, $n_s\in[0,\,1]$ is the solid fraction which comes from GS. It interpolates between fluidic (cf.\ lines~1 and 2 of \autoref{eq:LGS}) and solid contributions (cf.\ line 3~of \autoref{eq:LGS}) to $\boldsymbol{f}$ and can be used to describe homogenized regions, such as the binder. For $n_s=0$ or $n_s=1$, \autoref{eq:LGS} describes a purely fluid-like or solid-like behavior, respectively.

From the moments of $\boldsymbol{f}$, physical quantities such as density or momentum are determined (cf.\ \eqsref{eq:rho_BGK}{eq:mom_BGK}). Any fluid or partially fluid lattice cell is computationally simultaneously occupied by both components. Under most conditions, a cell consists of a main component with the bulk density $\rho$ and a dissolved component with the residual density $\rho_{\text{dis}} \ll \rho$. 

Three types of forces are modeled. Those are fluid-fluid interactions, solid-fluid interactions, and external forces. The fluid-fluid interaction force $\boldsymbol{F}^{\sigma}_\mathrm{inter}$ between the components $\sigma$ and $\bar{\sigma}$ is given by
\begin{align}
    \label{eq:SC_F_inter}
    \boldsymbol{F}^{\sigma}_\mathrm{inter}\left(\mathbf{x}\right) = -\rho^{\sigma}\left(\mathbf{x}\right) G^{\sigma\bar{\sigma}}_\mathrm{inter} \sum_{i}^{}w_i \rho^{\bar{\sigma}}\left(\mathbf{x}+\mathbf{c}_i \Delta t\right)\mathbf{c}_i \Delta t,
\end{align}
where $G^{\sigma\bar{\sigma}}_\mathrm{inter}$ is the interaction parameter that determines the strength of the cohesion, i.e. the interfacial tension. The lattice specific parameters $w_i$ and $\mathbf{c}_i$ denote the weights and lattice velocities, respectively. They are given in the appendix.

The solid-fluid interaction force $\boldsymbol{F}^{\sigma}_\mathrm{ads}$ which acts on the $\sigma$-component is
\begin{align}
    \label{eq:SC_F_ads}
    \boldsymbol{F}^{\sigma}_\mathrm{ads}\left(\mathbf{x}\right) = -\rho^{\sigma}\left(\mathbf{x}\right) G^{\sigma}_\mathrm{ads} \sum_{i}^{}w_i s\left(\mathbf{x}+\mathbf{c}_i \Delta t\right)\mathbf{c}_i \Delta t,
\end{align}
where $G^{\sigma}_\mathrm{ads}$ is the interaction parameter which determines the wetting behavior. For the original MCSC, i.e. for $n_s=0$, it is directly correlated with the contact angle $\theta$ \cite{Huang2007} (cf.\ \autoref{eq:contactAngle}). The function $s$ serves as indicator. Typically it is $s=1$ at solid cells and $s=0$ otherwise \cite{Martys1996}. In this work, following Pereira \cite{Pereira2016,Pereira2017,Pereira2019} it is $s = n_s^{\sigma}$.
    
The external force $\boldsymbol{F}_\mathrm{ext}$ contributing to each component is weighted by its density ratio
\begin{align}    
    \label{eq:SC_F_ext}
    \boldsymbol{F}^{\sigma}_\mathrm{ext} = \frac{\rho^{\sigma}}{\rho} F_\mathrm{ext},
\end{align}
where $\rho^{\sigma}$ can be determined as is given by \autoref{eq:rho_BGK} and $\rho=\sum_{\sigma}^{}\rho^{\sigma}$ is the total density.

The sum of the aforementioned force contributions (cf.\ \eqssref{eq:SC_F_inter}{eq:SC_F_ads}{eq:SC_F_ext}) determines the total force $\boldsymbol{F}^{\sigma}_\mathrm{tot}=  \boldsymbol{F}^{\sigma}_\mathrm{inter}+\boldsymbol{F}^{\sigma}_\mathrm{ads}+\boldsymbol{F}^{\sigma}_\mathrm{ext}$ acting on a lattice cell. Combining the Shan-Chen forcing approach \cite{Krueger2016} with GS, $\boldsymbol{F}^{\sigma}_\mathrm{tot}$ is finally incorporated as a force-induced contribution to the equilibrium velocity 
\begin{align}    
    \label{eq:GSSC_u_eq}
    \mathbf{u}^{\mathrm{eq},\sigma} = \frac{\sum_{\sigma}^{}\rho^{\sigma}\mathbf{u}^{\sigma}/\tau^{\sigma}}{\sum_{\sigma}^{}\rho^{\sigma}/\tau^{\sigma}} + \frac{\tau^{\sigma}\boldsymbol{F}^{\sigma}_{tot}}{\rho^{\sigma}}.
\end{align}

The equilibrium velocity $\mathbf{u}^{\mathrm{eq},\sigma}$ determines $\boldsymbol{f}^\mathrm{eq}$ (cf.\ \autoref{eq:Maxwell}). It must not be confused with the macroscopic streaming velocity $\mathbf{u}_\mathrm{macro}$ 
\begin{align}    
    \label{eq:GSSC_u_macro}
    \mathbf{u}_\mathrm{macro} =  \sum_{\sigma}^{}\frac{1-n_s^{\sigma}}{\rho}\left(\sum_{i}^{} f^{\sigma}_i \mathbf{c}_i + \frac{\boldsymbol{F}^{\sigma}_\mathrm{tot} \Delta t}{2} \right). 
\end{align}

In comparison to the model proposed by Pereira \cite{Pereira2016,Pereira2017,Pereira2019}, here the more common Shan-Chen forcing is used. Moreover, the redefinition of $\boldsymbol{F}^{\sigma}_\mathrm{tot}$ as described above cancels the scaling of the fluid-fluid and solid-fluid interaction forces in \autoref{eq:GSSC_u_eq} and, thus, maintains a thin interface. For the model adaptions, the effect of the solid fraction $n_s$ on the adhesive force $\boldsymbol{F}_\mathrm{ads}$ and the contact angle were determined. The results of which are given in the Supporting Information (cf.\ \autoref{SI-sec:ModelVerification}).

\subsection{Model Parametrization} \label{subsec:Parameters} 
In porous media applications gravitational and viscous forces are typically negligible compared to capillary or surface forces \cite{Weydanz2018,Danner2016,Chen2014,Li2018,Landry2014}. Thus, the right parametrization of density and viscosity ratios between two components has only a minor effect on the physical results of the simulation \cite{Li2018,Landry2014}. Therefore, and due to stability reasons of MCSC \cite{Chen2014} the density and viscosity ratio is set to unity. The other relevant model parameters and the corresponding conversion factors of the electrolyte-gas system studied here are given in  \autoref{tab:Parameters} and \autoref{tab:Conversion}, respectively. They are chosen to represent propylene carbonate as electrolyte and air as gas.

\begin{table}[h!]
	\centering
	\caption{Overview of the physical quantities of the system consisting of electrolyte (E) and gas (G). Values are given in SI units and LBM units (lu: length unit; ts: time step; mu: mass unit). The asterisk indicates quantities in SI units.}
	\begin{tabular}{l| l l}
		\toprule \toprule
                            & SI units                                      & LBM units \\
        \midrule
        density             & $\rho^{\mathrm{E,\ast}}\,=\,1.20\cdot10^{3}$ $\frac{\text{kg}}{\text{m³}}$ \cite{Moumouzias1992,Sun2018} & $\rho^{\mathrm{E}}\,=\,0.99$ $\frac{\text{mu}}{\text{lu³}}$ ($                          \rho_{\text{dis}}^{\mathrm{E}}\,=\,0.01$ $\frac{\text{mu}}{\text{lu³}}$)\\
                            & $\rho^{\mathrm{G,\ast}}\,=\,1.18$ $\frac{\text{kg}}{\text{m³}}$ \cite{Cengel2018}           & $\rho^{\mathrm{G}}\,=\,0.99$ $\frac{\text{mu}}{\text{lu³}}$ ($\rho_{\text{dis}}^{\mathrm{G}}\,=\,0.01$ $\frac{\text{mu}}{\text{lu³}}$)\\
        kin. viscosity      & $\nu^{\mathrm{E,\ast}}\,=\,2.314\cdot10^{-6}$ $\frac{\text{m²}}{\text{s}}$ \cite{Moumouzias1992} & $\nu^{\mathrm{E}}\,=\,1.667\cdot10^{-1}$ $\frac{\text{lu²}}{\text{ts}}$ \\
                            & $\nu^{\mathrm{G,\ast}}\,=\,1.57\cdot10^{-5}$ $\frac{\text{m²}}{\text{s}}$ \cite{Cengel2018} & $\nu^{\mathrm{G}}\,=\,1.667\cdot10^{-1}$ $\frac{\text{lu²}}{\text{ts}}$ \\
        surface tension     & $\gamma^{\ast}\,=\,4.10\cdot10^{-2}$ $\frac{\text{kg}}{\text{s²}}$ \cite{Sun2018} & $\gamma\,=\,7.68\cdot10^{-2}$ $\frac{\text{mu}}{\text{ts²}}$ \\ \hline
        simulation              &                   & $\tilde{\tau}^{\mathrm{E}}\,=\,\tilde{\tau}^{\mathrm{G}}\,=\,1.0$ \\
        parameters              &                   & $G^{\mathrm{EG}}_\mathrm{inter}\,=\,G^{\mathrm{GE}}_\mathrm{inter}\,=\,1.75$ \\
                                &                   & $G^{\mathrm{G}}_\mathrm{ads}\,=\,-G^{\mathrm{E}}_\mathrm{ads}\,=\,\frac{1}{4}G^{\mathrm{EG}}_\mathrm{inter}(\rho^{\mathrm{E}}-\rho_{\mathrm{dis}}^{\mathrm{E}})\cos\theta$\\
		\bottomrule \bottomrule 
	\end{tabular}%
	\label{tab:Parameters}%
\end{table}%

\begin{table}[h!]
	\centering
	\caption{Overview of the conversion factors between SI units and the corresponding LBM units (lu: length unit; ts: time step; mu: mass unit).}
	\begin{tabular}{l r||l r}
		\toprule \toprule
		length & $C_l\,=\,4.38\cdot10^{-7}$ $\frac{\text{m}}{\text{lu}}$ & time & $C_t\,=\,1.3818\cdot10^{-8}$ $\frac{\text{s}}{\text{ts}}$ \\
		mass & $C_m\,=\,1.0194\cdot10^{-16}$ $\frac{\text{kg}}{\text{mu}}$ & pressure & $C_p\,=\,1.2189\cdot10^{6}$ $\frac{\text{kg/m s²}}{\text{mu/lu ts²}}$\\
		kin. Viscosity & $C_{\nu}\,=\,1.3884\cdot10^{-5}$ $\frac{\text{m²/s}}{\text{lu²/ts}}$ & force density & $C_f\,=\,2.2941\cdot10^{9}$ $\frac{\text{m/s²}}{\text{lu/ts²}}$ \\
		\bottomrule \bottomrule 
	\end{tabular}%
	\label{tab:Conversion}%
\end{table}%

All simulations conducted for this study have been carried out with an extended version of the open-source LBM tool \textit{Palabos} (version 2.3) \cite{Latt2021}.

\section{Simulation Setup} \label{sec:Setup}

Artificially generated 3D lithium-ion battery cathode structures \cite{westhoff.2018} were used as a geometrical basis for all simulations. Some structures were additionally infiltrated with binder using a physically motivated algorithm as described in \cite{Hein2020}. Recall that the focus of the present paper is twofold. First, the pressure-saturation behavior during the filling is studied. It shows optimum pressure profiles that ensure a steady and uniform filling process. Second, the systems obtained at the end of the filling are analyzed. The analysis concerns the final electrolyte saturation, the size and spatial distribution of gas agglomerates being entrapped, and the permeability of electrolyte and gas in those partially saturated electrodes. Moreover, it is shown how an imperfect filling might affect the battery performance. 

\subsection{Artificial Generation of Electrode Structures} \label{subsec:StochStruct} 
The underlying cathode structures have been generated by means of the stochastic microstructure modeling framework which is described in \cite{westhoff.2018}. It consists of three steps. First, a force-biased collective rearrangement algorithm is used to model positions and sizes of active particles by a non-overlapping sphere packing \cite{moscinski.1989, bezrukov.2002}. The packing density corresponds to the predefined volume fraction of active material and is varied in the range $\phi_{\text{A}} = \{0.5,\,0.6,\,0.7\}$ to obtain different electrode densities. Note that the radii of the spheres are drawn from a Gamma distribution, denoted by $R_{\text{PS}}$, with some shape and rate parameters $\alpha,\beta>0$. The following three cases are considered: Small spheres ($\alpha = 3.94,\, \beta = 2.17\,\mu m^{-1}$), medium spheres  ($\alpha = 2.62,\, \beta = 1.05 \,\mu m^{-1}$), and large spheres ($\alpha = 2.65,\, \beta = 0.75 \,\mu m^{-1}$). The second modeling step involves a connectivity graph \cite{westhoff.2018} which is omitted here due to the high volume fractions of active material. The shape of the simulated particles follows
the distribution of particle shapes that is also observed in realistic electrode structures. Thus, in the third step, each sphere was replaced by a not necessarily spherical particle, i.e.\ a structural anisotropy is introduced which is small enough such that simulation results are only affected within the uncertainty of the method (cf.\ \autoref{SI-sec:UncertPressSat}). Particle sizes and shapes are described by means of a radius function. This function is represented by a truncated series expansion in terms of spherical harmonics \cite{feinauer.2015a}, with the truncation parameter $L=6$. The simulation of active particles represented in this way was carried out by means of Gaussian random fields on the sphere. The corresponding angular power spectrum is given by $a_{1}=0.65$, $a_{2}=4.13$, $a_{3}=0.82$, $a_{4}=0.31$, $a_{5}=0.17$, $a_{6}=0.11$, see \cite{feinauer.2015b} for details.

Finally, the system of simulated particles was discretized using a voxel size of $0.438 \,\mu m$. In dependence on the volume fraction $\phi_{\text{A}} = \{0.5,\,0.6,\,0.7\}$, the number of voxels was 82, 170, and \{388,\,323,\,277\} along the $x$-, $y$-, and $z$-direction, respectively. Note that periodic boundary conditions in $y$- and $z$-direction were applied in all simulations. For the simulation scenarios with the IDs~9$-$14, a volume fraction of $\phi_{\text{B}} = 0.21$ of the binder was added to the reference structure, i.e.\ ID~1 (cf.\ Table~\ref{tab:SimulationOverview}). Therefore, a morphological closing of the active material was applied where a sphere was used as the structuring element \cite{soille.2003}. The radius of the sphere was chosen such that the predefined volume fraction of the binder was obtained.

\subsection{Pressure-Saturation Behavior} \label{subsec:PressSat} 
\begin{figure}
	\centering
	  \includegraphics[width=0.8\textwidth]{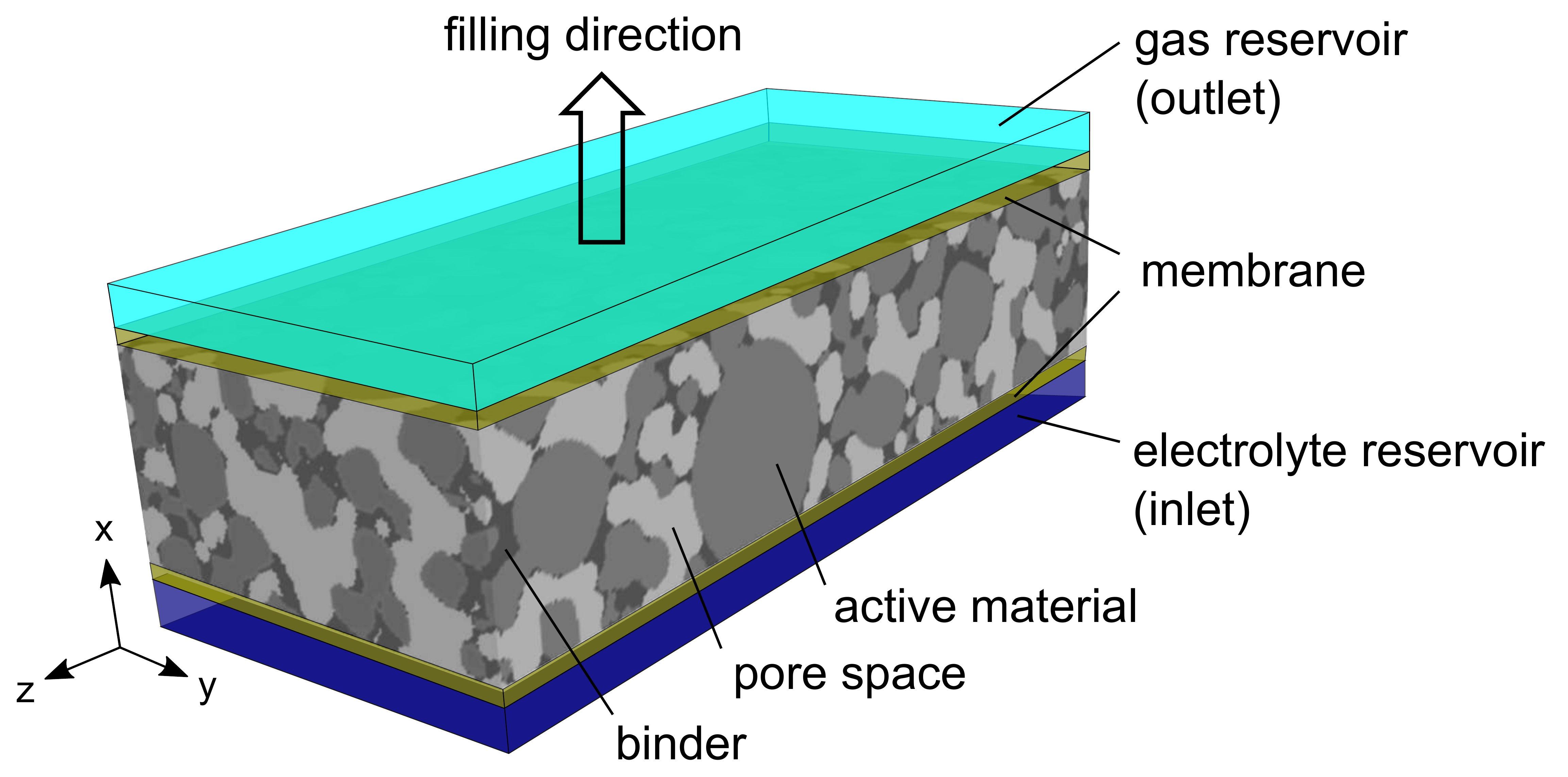}
	\caption{Scheme of the simulation setup. The electrode consists of active material (gray) and potentially a binder (darker gray) both enclosing the mesoscopic pore space (lighter gray). The reservoirs in which the densities of electrolyte and gas were prescribed are marked in blue and cyan, respectively. The membranes adjacent to the reservoirs are depicted in yellow. They were semi-permeable during the electrolyte filling process in $+x$-direction and impermeable during the permeability simulations in which a driving force was applied in $+y$-direction.}
	\label{fig:SystemSetup}
\end{figure}

The relationship between capillary pressure and saturation is an important measure for porous media applications. It is used to predict which capillary pressure has to be overcome to reach a certain saturation \cite{Falode2014,Akai2019}. Our simulation setup mimics experimental setups that are typically used to determine capillary pressure-saturation curves of porous media, e.g., in the context of fuel cells \cite{Gostick2008,Harkness2009,Fairweather2010,Dwenger2012}. A scheme of the simulation setup that was used to study the pressure-saturation behavior is shown in \autoref{fig:SystemSetup}. It consists of the electrode structure as described in \autoref{subsec:StochStruct}. The total pore space, i.e.\ the unification of the mesoscopic pores confined by the active material and the inner pores in the binder, was initially filled with a gas of density $\rho^{\mathrm{G}}$ and the dissolved electrolyte with density $\rho_{\text{dis}}^{\mathrm{E}}$ (cf.\ \autoref{tab:Parameters}). 

Large scale simulations have been conducted. For $\phi_{\text{A}} = \{0.5,\,0.6,\,0.7\}$, the system sizes were 40\,$\mu$m, 75\,$\mu$m, and \{170,\,140,\,120\}\,$\mu$m along the $x$-, $y$-, and $z$-direction, respectively. This corresponds to simulation domains of up to 5.9~million lattice cells. The simulations were computationally expensive and, thus, conducted on the supercomputers JUSTUS~2 and Hawk using more than 500~cores in parallel execution.

Periodic boundary conditions were applied along the $y$- and $z$-direction. Along the $x$-direction an electrolyte reservoir and a gas reservoir were added at the inlet and outlet, respectively. The reservoirs had a thickness of four layers each. They were used to prescribe the density of the corresponding fluid, i.e.\ proportional to the pressure (cf.\ \autoref{eq:SC_pressure}). The initial electrolyte density at the inlet was $\rho^{\mathrm{E}}$ (and the gas density to $\rho_{\text{dis}}^{\mathrm{G}}$). During the simulation run, $\rho^{\mathrm{E}}$ at the inlet was incrementally increased using a control loop under the condition of steady and uniform filling with a predefined target saturation rate. The gas density at the outlet was constant, i.e.\ $\rho^{\mathrm{G}}$ (and the electrolyte density at $\rho_{\text{dis}}^{\mathrm{E}}$). Thereby, a pressure difference between the two fluid phases is applied that corresponds to the capillary pressure (cf. \autoref{eq:CapPressure}). This approach is in accordance with experiments and simulations, where the capillary pressure is adjusted by increasing or decreasing the pressure of the wetting phase or the nonwetting phase, respectively \cite{Gostick2008,Harkness2009,Fairweather2010,Dwenger2012,Satjaritanun2017,Niu2018,Zhu2021}. Each reservoir was divided from the electrode by a semi-permeable membrane to prevent an unwanted fluid breakthrough. The inlet membrane was permeable for the electrolyte only. The outlet membrane was permeable for the gas only. The impermeability was implemented by applying the bounce-back scheme (cf.\ \cite{Krueger2016}) to the non-permeating fluid.

From the simulations the pressure difference $\Delta p$ was determined as
\begin{align}    
    \label{eq:CapPressure}
    \Delta p = \left<p\right>_{\text{inlet}} - \left<p\right>_{\text{outlet}},
\end{align}
where $p$ was evaluated using \autoref{eq:SC_pressure}, and $\left<p\right>$ denotes the average pressure in the inlet and outlet reservoirs. The pressure difference $\Delta p$ is directly related to the capillary pressure $p_c$ as $p_c = p_0 -\Delta p$, where $p_0$ is the absolute capillary pressure at zero electrolyte saturation ($p_0 = p_c(S^{\mathrm{E}}=0)$). In the current work, $\Delta p$ was chosen over $p_c$ to improve the comparability and ensure that all pressure-saturation curves start from the same value $\Delta p(S^{\mathrm{E}}=0) = 0$.

The electrolyte saturation $S^{\mathrm{E}}$ is defined as
\begin{align}    
    \label{eq:Saturation}
    S^{\mathrm{E}} = \frac{N_{\text{pore}}(\rho^{\mathrm{E}}\geq0.5)+(1-n_s)N_{\text{binder}}(\rho^{\mathrm{E}}\geq0.5)}{N_{\text{pore}}+(1-n_s)N_{\text{binder}}},
\end{align}
where the denominator and numerator correspond to the total pore space and the pore space in which $\rho^{\mathrm{E}}\geq0.5$ mu/lu$^3$, respectively. The number of pore lattice cells in the electrode structures and the binder are denoted by $N_{\text{pore}}$ and $N_{\text{binder}}$, respectively. The latter are multiplied by the effective nanoscopic pore volume $(1-n_s)=(1-\phi'_{\text{B}})$. Note, that for the calculation of the saturation only the lattice cells between the two membranes were considered.

A simulation run consisted of approximately 1,000,000 time steps. Only the two simulations in which the process time was varied by the factor 0.5 and 2, accordingly consisted of approximately 500,000 and 2,000,000 time steps. The pressure difference and the saturation were determined every 10,000 time steps during the production run. The simulations were stopped when a further saturation was not possible and led to a steep increase of $\Delta p$. The corresponding distribution functions of both fluids were used for further data analysis and as input for subsequent permeability simulations.

The statistical uncertainty of the pressure-saturation curves was estimated for a representative electrode structure. The mean standard deviation of the average curve was 2$-$4\,kPa for $\Delta p$ and to 0.75\,\% for $S^{\mathrm{E}}$. Detailed results are given in the Supporting Information (cf.\ \autoref{SI-sec:UncertPressSat}). 

\subsection{Gas Entrapment} \label{subsec:GasBubbleAnalysis} 
As was recently reported by Sauter~\textit{et al.}\ \cite{Sauter2020}, gas entrapment can significantly reduce effective ionic conductivities in separators. The gas phase is a poor conductor that hinders ion transport, blocks transport pathways, and reduces the connectivity of the electrolyte phase. This can be quantified by the mean geodesic tortuosity \cite{stenzel.2016, neumann.2020}. It is determined by the lengths of shortest paths between inlet and outlet plane completely contained in a predefined phase. In this study, it is computed using Dijkstra's algorithm \cite{jungnickel.2013}. By dividing the lengths of those shortest paths by the thickness of the electrode in $x$-direction and by subsequent averaging over all starting points in the inlet, an estimator of the mean geodesic tortuosity is obtained. For a more formal introduction to geodesic tortuosity, see \cite{neumann.2019a}. 

Two different geodesic tortuosities are determined, i.e. $\tau_{0}$ and $\tau_{\text{end}}$. For $\tau_{0}$, the active material and partially also the binder are considered as obstacles for the ionic transport. Thus, $\tau_{0}$ represents the geodesic tortuosity for an ideal case in which each pore contributes to the ionic pathways. In contrast, $\tau_{\text{end}}$ is the geodesic tortuosity at the end of the filling process. Then, also entrapped gas is an obstacle for densities exceeding $\rho^{\mathrm{G}}\geq0.5$. In case of simulations with binder (IDs 9-14), an additional weighting factor $w_{\text{B}}$ accounts for increased path lengths within a binder. More precisely, the equation $w_{\text{B}} = \phi_{\text{B}}^{\prime -0.5}$ is used, which corresponds to the Bruggeman relation \cite{bruggeman.1935} and the frequently used Bruggeman exponent of -0.5 \cite{vadakkepatt.2015, patel.2003}. 

In addition, gas that accumulates at the surface of active material reduces the electrochemically active surface area and, thereby, limits the lithiation process. Thus, blocked surface areas of active material are analyzed. For this purpose, the surface area of active material ($S_{\text{A}}$), gas ($S_{\text{G}}$) and the union of both ($S_{\mathrm{A\cup G}}$) is estimated from voxelized image data by means of differently weighted local $2\times 2 \times 2$ voxel configurations, using the weights proposed in \cite{schladitz.2007}. Thereby, the fraction $S_{\mathrm{A\cap G}}$ of blocked active material surface is determined as
\begin{equation}
\label{eq:surfaceArea}
    S_{\mathrm{A \cap G}} = \frac{S_{\text{A}} + S_{\text{G}} - S_{\mathrm{A\cup G}}}{2S_{\text{A}}} \;\in [0,\,1].
\end{equation}
Note that the interfacial area between active material and gas contributes to $S_{\text{A}}$ and $S_{\text{G}}$ but not to $S_{\mathrm{A\cup G}}$, which leads to the factor of 2 in the denominator. 

\subsection{Permeability} \label{subsec:Perm} 
The permeability $k$ is a measure for the ability of a porous medium to perfuse fluid flow. Thus, it represents fluid mobility. In this study, the permeability is used to quantify the effort that is necessary for displacing entrapped gas agglomerates from electrodes.

The simulation setup for determining the permeability is similar to the setup in \autoref{subsec:PressSat}. Only deviations from this setup are described here. Electrolyte and gas distributions were initialized identical to those at the end of the filling process. Periodic boundary conditions were applied along all directions. The membranes were fully impermeable to conserve the fluid composition within the electrode. The densities of both fluids were constant. Along the positive $y$-direction the external force density $\mathsf{f}_y = 5\cdot10^{-4}\, \text{lu\,ts}^{-2}$ was applied. It was chosen such that the momentum showed a linear relationship with the external force \cite{Landry2014,Li2005,Martys1996}. The permeabilities were determined between the two membranes only.

From the simulations the permeability $k^{\sigma}_y$ of the component $\sigma$ along the $y$-direction was determined as
\begin{align}    
    \label{eq:Permeability}
    k^{\sigma}_y = \frac{u_{\text{Darcy},y}^{\sigma}\nu^{\sigma}}{\mathsf{f}_y}.
\end{align}

Thus, $k^{\mathrm{E}}_y$ and $k^{\mathrm{G}}_y$ denote the permeabilities of the electrolyte and the residual gas phase, respectively. While $\nu$ and $\mathsf{f}_y$ were input parameters to the simulations, the Darcy velocity $u_{\text{Darcy},y}^{\sigma}$ of the $\sigma$-component in $y$-direction was determined from the simulations as
\begin{align}    
    \label{eq:vDarcy}
    u_{\text{Darcy},y}^{\sigma} = \frac{\sum_{j}^{N_{\sigma,\text{bulk}}}u_y^{\sigma}(x_j)}{N_{\text{total}}}.
\end{align}
The Darcy velocity is the sum of the velocity component in the $y$-direction, $u_y$, over all lattice cells $j\in N_{\sigma,\text{bulk}}$ that belong to the bulk phase of the $\sigma$-component $N_{\sigma,\text{bulk}}$, divided by the total number of lattice cells $N_{\text{total}}$. The bulk phase did only contain fluid lattice cells without direct contact to a solid and in which the dissolved density of the complementary $\bar{\sigma}$-component, $\rho_{\text{dis}}^{\bar{\sigma}}$, was below 0.2\,mu\,lu$^{-3}$ to reduce errors from spurious currents \cite{Chen2014,Li2018}.

Each simulation consisted of two parts. Initially 100,000 time steps were performed in which a steady state was established. Subsequently, a production run of 100,000 time steps was conducted during which the permeability was determined every 1,000 time steps.

\section{Overview of the Study} \label{sec:StudyOverview} 

The influencing factors on the electrolyte filling process that are considered in this study are: the filling speed which corresponds to the process time $t_{\text{P}}$, the particle size distribution $R_{\text{PS}}$, the volume fraction $\phi_{\text{A}}$, and the wettability $\theta_{\text{A}}$ of active material, as well as the inner volume fraction $\phi'_{\text{B}}$ and wettability $\theta_{\text{B}}$ of binder. The volume fraction $\phi_{\text{A}}$ is the number of active material lattice cells divided by the total number of lattice cells. In contrast, the inner volume fraction $\phi'_{\text{B}}$ is the volume fraction of solid binder within a single binder lattice cell. Here it was assumed that $\phi'_{\text{B}} = n_s$, which is a simplification and not generally true \cite{Walsh2009,Pereira2019}.

An overview of the simulations from the current work is given in \autoref{tab:SimulationOverview}. To determine the pressure-saturation behavior, 16 large-scale 3D simulations were conducted. Another 16 simulations were conducted to determine permeabilities. The material properties of active material (IDs~1$-$8) and binder (IDs~9$-$14) were varied in a range that represents relevant electrode structures. Also the contact angles on the active material ($\theta_{\text{A}}=[60,100]$°) and the binder ($\theta_{\text{B}}=[30,120]$°) were chosen in a range that was observed in experiments \cite{Wu2004,Sun2018,Rosman2019,Sauter2020}. For all aforementioned simulations, i.e.\ IDs~1$-$14, the target saturation rate was identical. The IDs~15 and 16 refer to simulations in which the process time was varied.

Two simulations were used as reference, where all parameters were chosen such that they were in the middle of the parametric ranges studied in this work. Unless specified otherwise, subsequent simulations use those default parameters. The default simulation without binder is denoted as ID~1 (cf.\ first column in \autoref{tab:SimulationOverview}). The default simulation with binder is denoted as ID~9.

\begin{table}[h!]
	\centering
	\caption{Overview of the performed simulations. The default simulation without binder is ID~1. The default simulation with binder is ID~9, where the properties of the active material were identical to those of ID~1. The simulations 1$-$8, 15, and 16 did not contain any binder and thus are marked by '$-$' in columns 5 and 6. In contrast, the simulations 9$-$14 contained binder. Plain entries of settings have the same value as ID~1 or 9, respectively. Beside the total duration of the filling process $t_{\text{end}}$, also the results of the final saturation $S^{\mathrm{E}}_{\text{final}}=S^{\mathrm{E}}(t_{\text{end}})$, as well as the permeability of the electrolyte $k^{\mathrm{E}}_y$ and the residual gas phase $k^{\mathrm{G}}_y$ are given. }
	\begin{tabular}{c|ccc|cc|cccc}
		\toprule \toprule
        sim. ID   & $R_{\text{PS}}$ & $\phi_{\text{A}}$ & $\theta_{\text{A}}$ & $\phi'_{\text{B}}$ & $\theta_{\text{B}}$ & $t_{\text{end}}$ & $S^{\mathrm{E}}_{\text{final}}$ & $k^{\mathrm{E}}_y$ & $k^{\mathrm{G}}_y$ \\
             &                 &      & (°)        &      & (°)        & ($10^{-2}$ s)    & (\%)                & ($10^{-15}$ m²) & ($10^{-15}$ m²) \\
		\midrule
	 	1    & medium    & 0.6 & 90 & $-$  & $-$  & 1.46 & 89.6 &  91.72 &  9.62 \\
        2    & small     &     &    & $-$  & $-$  & 1.41 & 90.7 & 102.58 & 10.55 \\
	 	3    & large     &     &    & $-$  & $-$  & 1.52 & 95.3 & 198.66 & 70.64 \\
        4    &           & 0.7 &    & $-$  & $-$  & 1.26 & 79.1 &  32.28 &  5.82 \\
	 	5    &           & 0.5 &    & $-$  & $-$  & 1.48 & 95.6 & 264.33 & 97.59 \\
        6    &           &     & 60 & $-$  & $-$  & 1.48 & 96.8 &  88.98 & 40.97 \\
	 	7    &           &     & 80 & $-$  & $-$  & 1.48 & 92.7 &  92.16 & 16.38 \\
        8    &           &     &100 & $-$  & $-$  & 1.42 & 86.1 &  91.64 &  9.66 \\ \hline
	 	9    & medium    & 0.6 & 90 & 0.5  &  60  & 1.39 & 95.9 &   9.82 &  1.00 \\
        10   &           &     &    & 0.6  &      & 1.39 & 95.5 &   7.21 &  1.06 \\
        11   &           &     &    & 0.4  &      & 1.38 & 95.3 &  13.19 &  1.64 \\
        12   &           &     &    &      &  30  & 1.35 & 96.6 &  10.03 &  5.93 \\
        13   &           &     &    &      &  90  & 1.35 & 89.1 &   8.80 &  1.75 \\
        14   &           &     &    &      & 120  & 1.02 & 63.0 &   4.99 &  1.89 \\ \hline
        15   & medium    & 0.6 & 90 & $-$  & $-$  & 2.98 & 90.1 &  92.21 &  8.11 \\
        16   & medium    & 0.6 & 90 & $-$  & $-$  & 0.73 & 89.3 &  91.30 & 10.63 \\
		\bottomrule \bottomrule
	\end{tabular}%
	\label{tab:SimulationOverview}%
\end{table}%

Starting from ID~1, all influencing factors concerning the active material were studied independently. The parameters $R_{\text{PS}}$, $\phi_{\text{A}}$, and $\theta_{\text{A}}$ were varied separately, while the other influencing factors were kept constant at their default values. The influence of the binder was studied by separately varying $\phi'_{\text{B}}$ or $\theta_{\text{B}}$, while keeping all other influencing factors constant at the values identical to those from ID~1. The structural properties of the IDs~15 and 16 were identical to those of ID~1. 

Filling and permeability simulations were conducted for each ID. Numerical results are also given in \autoref{tab:SimulationOverview}.

\section{Results and Discussion} \label{sec:Results} 

\subsection{Pressure-Saturation Behavior} \label{subsec:PressResults} 
\begin{figure}
	\centering
	  \includegraphics[width=0.96\textwidth]{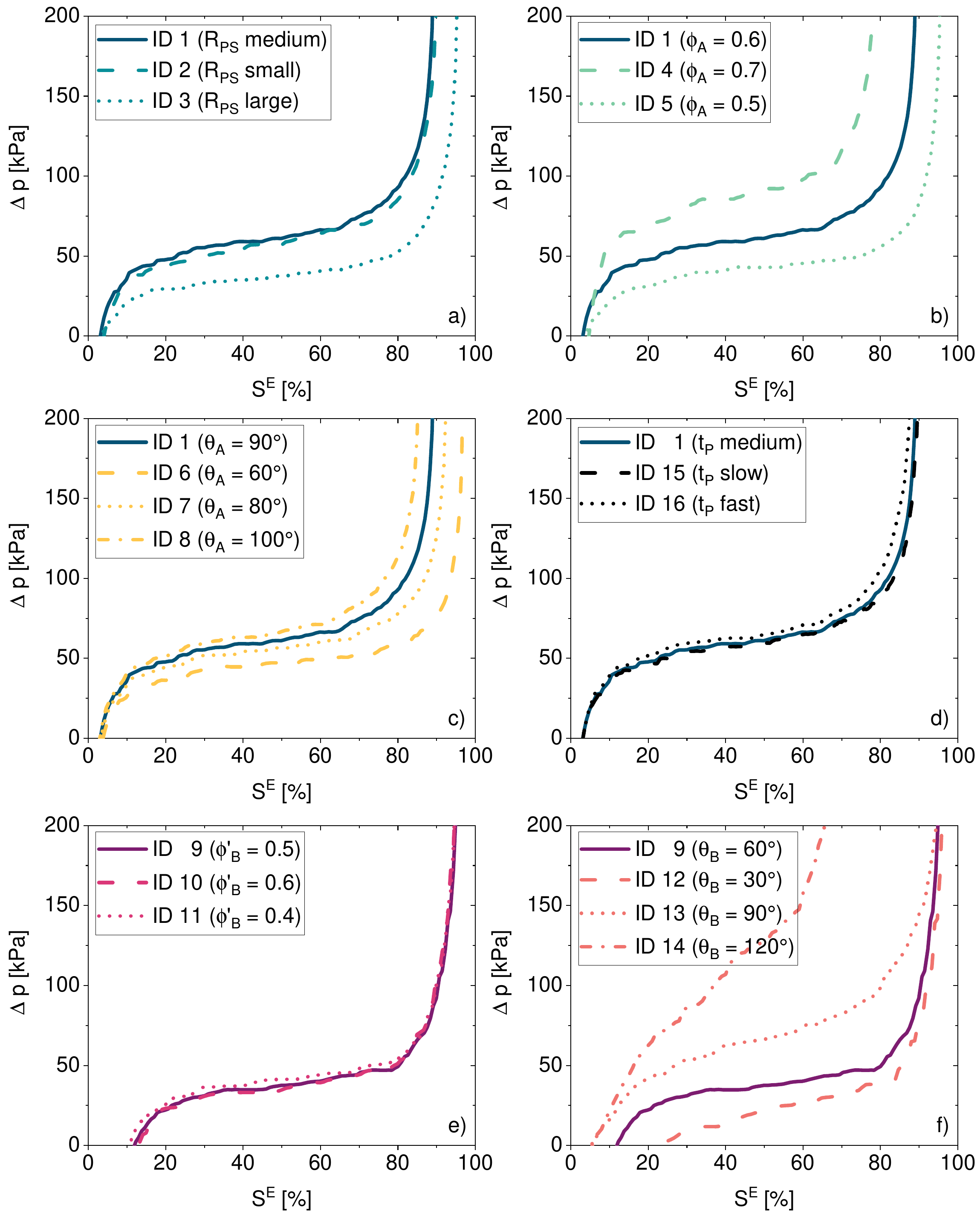}
	\caption{Pressure-saturation behavior of electrodes without, i.e.\ a)$-$d), and with binder, i.e.\ e)$-$f). ID~1 and ID~9 are depicted with the blue and purple solid lines, respectively. The influencing factors are indicated by the colors. Those are a) the particle size distribution $R_{\text{PS}}$ (turquoise), b) the volume fraction of the active material $\phi_{\text{A}}$ (green), c) the wettability of the active material $\theta_{\text{A}}$ (orange), d) the process time $t_{\text{P}}$ (black), e) the inner volume fraction of the binder $\phi'_{\text{B}}$ (magenta), and f) the wettability of the binder $\theta_{\text{B}}$ (red). The line types correspond to a specific value of the influencing factor.}
	\label{fig:pS_total_all}
\end{figure}

\autoref{fig:pS_total_all} shows the pressure-saturation curves of all simulations. They follow a sigmoidal behavior with steep sides for low and high saturations, and an almost horizontal regime for medium saturations. This trend can be explained by the Young–Laplace equation ($p_c=2\gamma/R$) which describes the inverse proportionality between the capillary pressure $p_c$ and the pore radius $R$. When electrolyte initially invades the electrode, smaller pores at the inlet need to get passed leading to an increase in $\Delta p$. Thereafter, a plateau is reached, during which the electrodes are primarily filled through larger pores. Finally, for high saturations, smaller pores have to be filled, leading to a strong increase in $\Delta p$ again. For all cases, the final saturation $S^{\mathrm{E}}_{\text{final}}$ deviates from the theoretical optimum of 100\,\% which is related to gas agglomerates being entrapped in the pore space  \cite{Schaap2007,Knoche2016,Liu2016,Guenter2018,Weydanz2018,Li2018,Avendano2019,Sauter2020,Shodiev2021}.

\autoref{fig:pS_total_all}\,a)\,$-$\,d) show the results for the influencing factors that are related to the active material and the process time. \autoref{fig:pS_total_all}\,e)\,\&\,f) show the results purely related to the binder. There and in all figures in the following, the influencing factors are indicated by the colors. The line types correspond to a specific value of the influencing factor. In \autoref{fig:pS_total_all}, the results of the reference cases ID~1 and ID~9 are depicted by the blue and purple solid lines, respectively.

\autoref{fig:pS_total_all}\,a)\,\&\,b) show the influence of the particle size distribution $R_{\text{PS}}$ and the volume fraction $\phi_{\text{A}}$ of the active material. Compared to the reference, larger particle sizes (ID~3) and a smaller volume fraction of the active material (ID~5) result in a smaller $\Delta p$ and an increased final saturation $S^{\mathrm{E}}_{\text{final}}$. Both are related to larger pores and reduced $\Delta p$. The contrary is observed for larger $\phi_{\text{A}}$ (ID~4) which facilitates gas entrapping.

The influence of the wetting behavior of active material is shown in \autoref{fig:pS_total_all}\,c). The results indicate that decreasing $\theta_{\text{A}}$ or increasing the wettability reduces $\Delta p$ and improves the saturation. 

\autoref{fig:pS_total_all}\,d) shows that there is hardly any influence of the process time $t_{\text{P}}$ for the values studied here. The medium (ID~1) and slow (ID~15) filling processes were slow enough such that capillary forces dominated viscous forces. For a fast filling (ID~16), viscous effects are more apparent \cite{Huang2014,Li2018}. The flow regime then tends to transition from capillary fingering to viscous fingering which leads to more gas entrapment \cite{Lenormand1988,Huang2014}. 

\autoref{fig:pS_total_all}\,e)\,\&\,f) show the influence of the binder. In general, the binder shifted $S^{\mathrm{E}}_{\text{init}}$ to larger values. This was partially due to the definition of the saturation (cf.\ \autoref{eq:Saturation}), where adding a solid binder reduces the total pore space and was even more pronounced for strong wettabilities.

There is almost no influence of the inner volume fraction of the binder for the values studied here. This is different for the binder wettability. Using a strongly wetting binder (IDs~9$-$12) decreases $\Delta p$, enhances the electrolyte percolation, increases $S^{\mathrm{E}}_{\text{init}}$, and improves the final saturation. In contrast, using a neutrally wetting (ID~13) or dewetting binder (ID~14) causes larger $\Delta p$. Moreover, a dewetting binder leads to effects similar to pore clogging. It prevents electrolyte invading the binder and entraps large amounts of residual gas in the binder and at its surface.

\subsection{Total Duration of the Filling Process} \label{subsec:Duration} 
The saturation-time behavior for different process times or target saturation rates is given in \autoref{SI-fig:SaturationTime} in the Supporting Information. The results show a similar qualitative behavior. The control function adjusts the inlet density increment such that a steady and uniform filling process is achieved. This is shown by the almost linear behavior of the saturation-time curves. Deviations from that behavior occur at the beginning and the end of the simulations, where $\Delta p$ is highly sensitive to the saturation (cf.\ \autoref{fig:pS_total_all}).

As both a fast filling and a low final saturation decrease $t_{\text{end}}$, a relative measure, i.e.\ the reciprocal filling rate
\begin{align}    
    \label{eq:relDuration}
    \tilde{t}_{\text{end}}=\frac{t_{\text{end}}}{S^E_{\text{final}}},
\end{align}
is introduced. It corresponds to the average time needed to fill 1\,\% of  electrode's pore space.

\begin{figure}
	\centering
	  \includegraphics[width=0.6\textwidth]{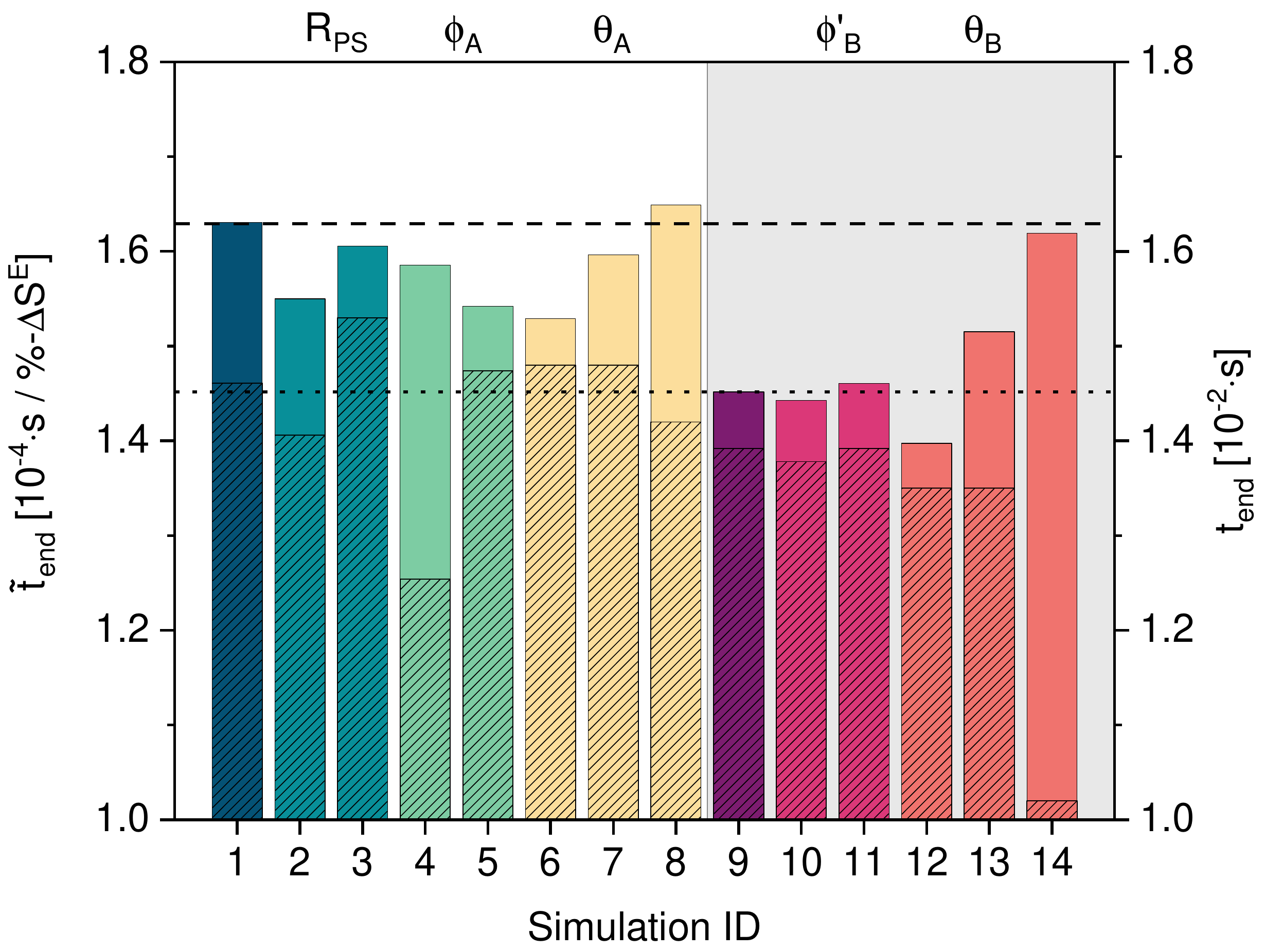}
	\caption{Overview of the reciprocal filling rate $\tilde{t}_{\text{end}}$ and the total duration of the filling process $t_{\text{end}}$. The values of $\tilde{t}_{\text{end}}$ are depicted with the colored bars and related to the left ordinate axis. The values of $t_{\text{end}}$ are depicted with the hatched bars and related to the right ordinate axis. The influencing factors are indicated by the colors. The corresponding simulation IDs are given at the abscissa. Simulations of electrodes with binder are highlighted by the gray background. The horizontal dashed and dotted lines represent $\tilde{t}_{\text{end}}$ of ID~1 and ID~9, respectively.}
	\label{fig:Duration}
\end{figure}

For the IDs~1$-$14, the results of $t_{\text{end}}$ and $\tilde{t}_{\text{end}}$ are shown in \autoref{fig:Duration}. There is a clear correlation between $t_{\text{end}}$ and the structural properties of the active material (IDs~1$-$5). The smaller the pores are, the shorter is the total duration. Moreover, there is a strong dependence between $\tilde{t}_{\text{end}}$ and $\theta_{\text{A}}$. Stronger wettabilities result in lower reciprocal filling rates and, thus, shorter filling processes. The same effect is observed for the binder wettability (IDs~9, 12$-$14) and has also been reported in the literature \cite{Wu2004,Lee2013,Wolf2020}. In general, the filling of electrodes with binder is about 20\,\% faster compared to electrodes without binder. However, this is also related to the reduction of the total pore space when adding binder.

\subsection{Final Saturation and Gas Entrapment} \label{subsec:GasResults}
\begin{figure}
	\centering
	  \includegraphics[width=0.6\textwidth]{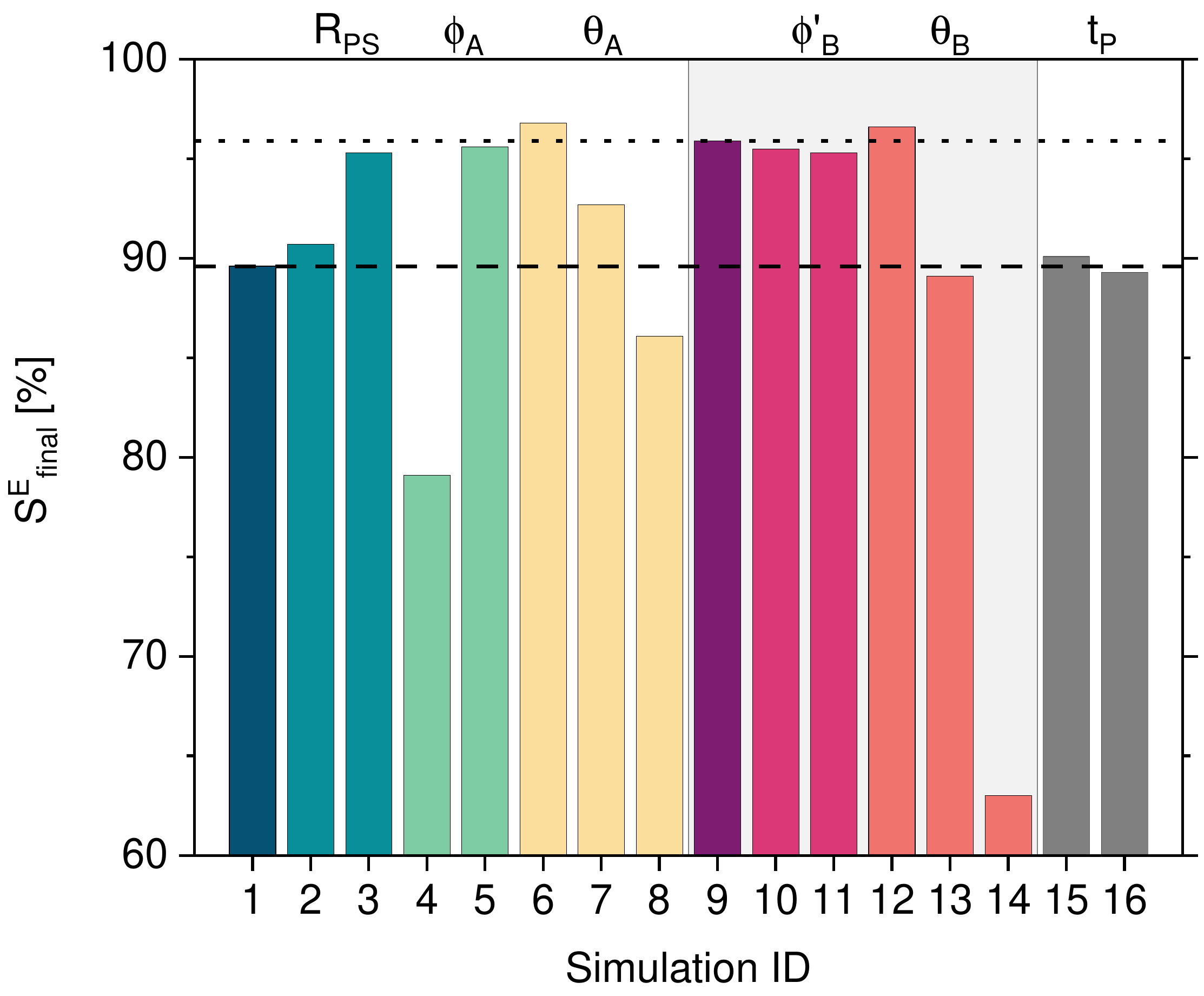}
	\caption{Overview of the final saturation of the filling process $S^E_{\text{final}}$. The meaning of the colors is identical to those from \autoref{fig:Duration}. The horizontal dashed and dotted lines represent $S^{\mathrm{E}}_{\text{final}}$ of ID~1 and ID~9, respectively. }
	\label{fig:Saturation}
\end{figure}

First, the final saturations at the end of the filling process are compared. The numerical values of $S^{\mathrm{E}}_{\text{final}}$ are listed in \autoref{tab:SimulationOverview} and shown in \autoref{fig:Saturation}. 

The final saturations are in a broad range $S^{\mathrm{E}}_{\text{final}}=[63.0,\,96.8]\,\%$ which corresponds to a residual gas volume fraction of $[37.0,\,3.2]\,\%$. Most of the electrodes are filled 90\,\% and more, which is in accordance with observations from experiments \cite{Weydanz2018}. Compared to ID~1 a larger saturation is observed for electrodes with 1)~larger pores, i.e.\ larger $R_{\text{PS}}$ (ID~3) or smaller $\phi_{\text{A}}$ (ID~5), 2)~better wettability (IDs~6 and 7), and 3)~in the presence of a hydrophilic binder (IDs~9$-$12). In contrast, incomplete filling correlates with 1)~small pores (ID~4) and 2)~hydrophobic active material (ID~8) and binder (ID~14). These general findings have been shown for single influencing factors in experimental \cite{Wu2004,Habedank2019,Schilling2019,Akai2019,Sauter2020,Guenter2022} and simulative \cite{Lee2013,Mohammadian2018,Jeon2019,Akai2019,Wolf2020,Shodiev2021,Shodiev2021ML} studies in the literature. Here, they are quantified and summarized for a broad variety of decoupled influencing factors. Together with the high spatial resolution of LBM in the sub-micrometer range and a detailed analysis, the results of the present work go far beyond the state-of-the-art knowledge and are further discussed in the following. 

\begin{figure}
  \centering
      \includegraphics[width=1\textwidth]{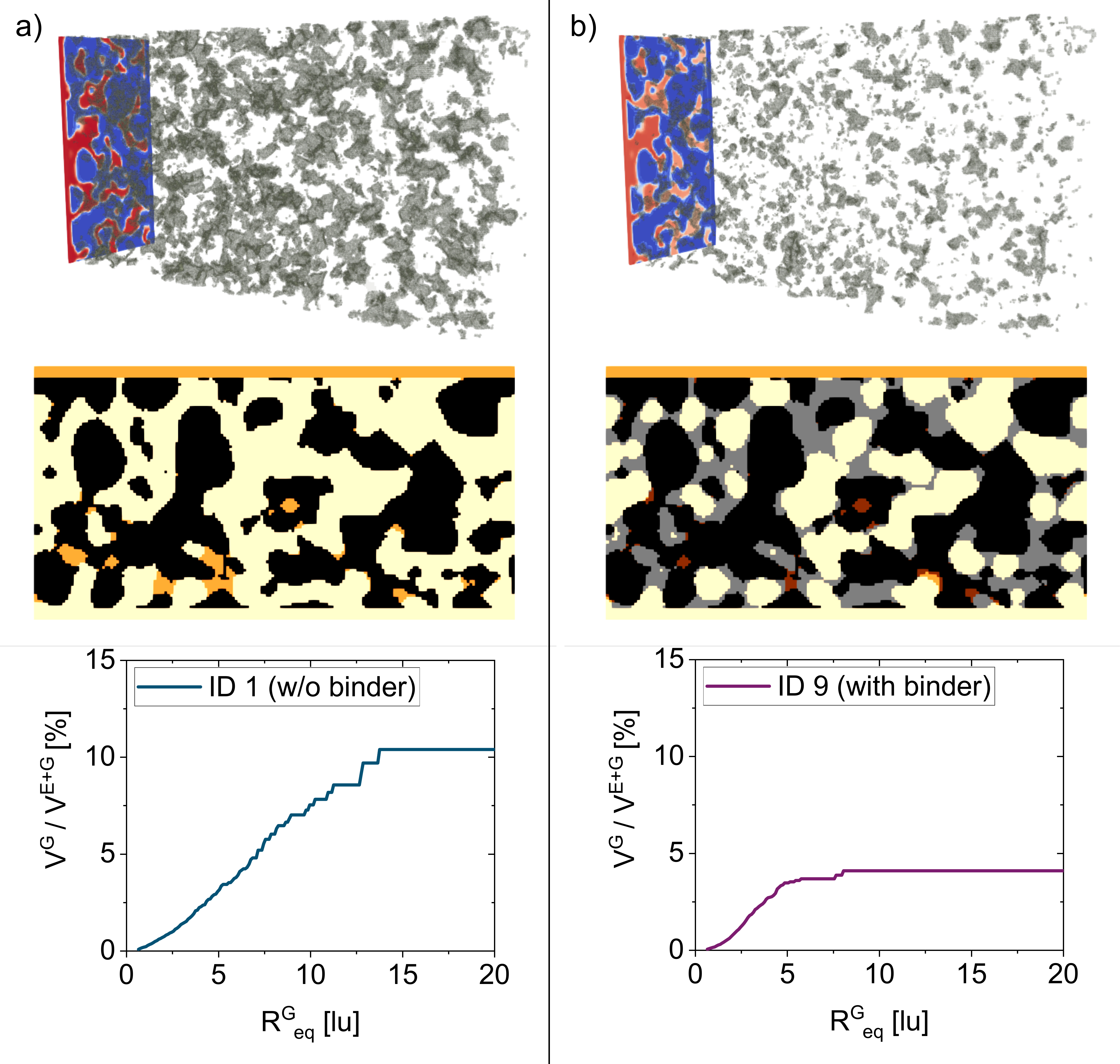}
    \caption{Comparison of the gas entrapment for the two reference simulations: a) ID~1 with $10.4\,\%$ residual gas phase and b) ID~9 with $4.1\,\%$ residual gas phase. Top: Visualization of the residual gas phase being entrapped in the electrodes at the end of the filling process. The gas phase is depicted in gray. The electrolyte and the solid components are fully transparent. Middle: Cross section through the $xy$-plane at $z=200\,\text{lu}$. The active material is depicted in black, the binder is depicted in gray, and the gas phase is depicted in red or orange in regions with or without binder, respectively. Bottom: Size distributions of gas agglomerates. The ratio of the cumulated gas volume $V^{\text{G}}$ to the total pore volume $V^{\text{E+G}}$ is shown as a function of the equivalent gas bubble radius $R^{\text{G}}_{\text{eq}}$. }
  \label{fig:GasBubbles}
\end{figure}

\autoref{fig:GasBubbles} shows qualitative and quantitative information of the gas agglomerates for the two reference cases. The amount of residual gas phase is $10.4\,\%$ in \autoref{fig:GasBubbles}\,a) and $4.1\,\%$ in \autoref{fig:GasBubbles}\,b). In the top figure, the gas entrapment is shown qualitatively. There, gas agglomerates are depicted in gray. All other components are fully transparent. In the middle figure, a cross section through the $xy$-plane at $z=200\,\text{lu}$ is shown. There, active material, binder, and gas phase are depicted in black, gray, and red or orange in regions with or without binder, respectively. In the bottom figure, the corresponding size distributions of the gas agglomerates are given. They show the ratio of cumulated gas volume $V^{\text{G}}$ to total pore volume $V^{\text{E+G}}$ as a function of the equivalent gas bubble radius $R^{\text{G}}_{\text{eq}}$. 

The top row of \autoref{fig:GasBubbles} shows that more gas phase and larger agglomerates are entrapped for ID~1, especially close to the inlet. This is also observed from the middle row of \autoref{fig:GasBubbles} where most of the gas agglomerates are in the lower half of the figure. For ID~1, gas agglomerates are mainly entrapped in small pores and corners confined by active material. The location of gas agglomerates is similar to ID~9. However, since the binder has a better wettability than the active material gas agglomerates are smaller as is also shown in the quantitative plots at the bottom row of \autoref{fig:GasBubbles}. 

The corresponding size distributions of all simulations are given in the Supporting Information (cf.\ \figsref{SI-fig:cumulatedVolume_noBinder}{SI-fig:cumulatedVolume_Binder}). The most relevant findings of which are summarized in the following: 1)~For almost all influencing factors, the slope $\Delta \left( V^{\text{G}}/V^{\text{E+G}} \right) / \Delta R^{\text{G}}_{\text{eq}}$ is similar until the asymptotic end value of $V^{\text{G}}/V^{\text{E+G}}$ is approached. This indicates a similar gas entrapment for small and medium gas agglomerates amongst all structures. 2)~Larger volume fractions of active material and smaller pores lead to larger gas agglomerates and better connectivity of the gas phase. 3)~A strong wettability of both active material and binder reduces gas entrapment and the size of gas agglomerates. 4)~The process time slightly affects the size distribution of gas agglomerates. Applying a fast filling speed (ID~16) leads to the formation of more medium-sized gas agglomerates.

The residual gas phase with its low ionic conductivity, is known to have a twofold impact on the battery performance \cite{Knoche2016,Knoche2016CIRP,Weydanz2018,Schilling2020,Sauter2020,Shodiev2021}. Gas agglomerates inhibit the ion transport, leading to longer transport pathways, and thereby decreasing the effective ionic conductivity. In addition,  gas prevents ion transport to the surface of the active material, reduces its electrochemically active surface area, increases overpotentials, and reduces the specific battery capacity. 

\begin{figure}
	\centering
	  \includegraphics[width=0.7\textwidth]{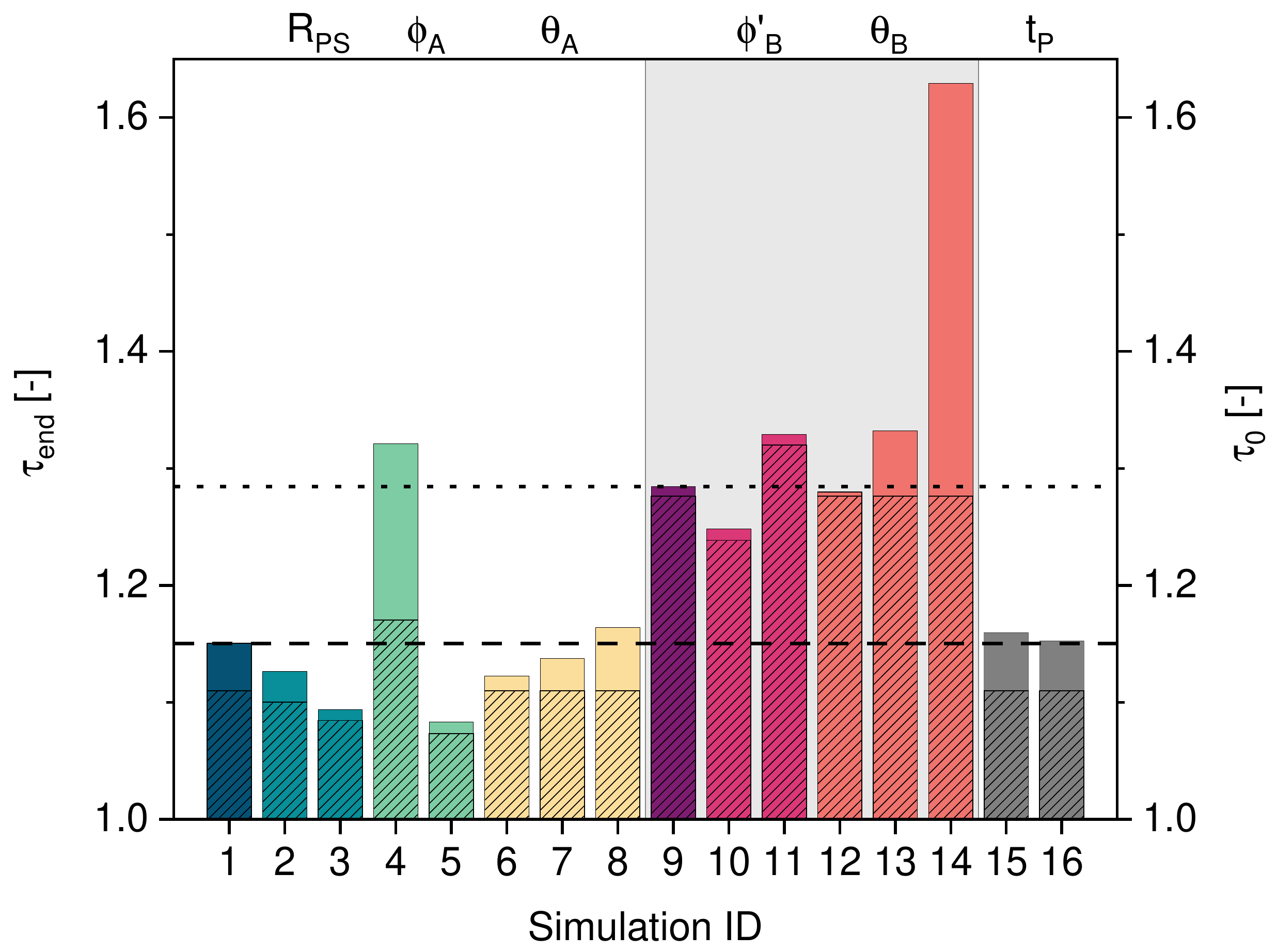}
	\caption{Overview of geodesic tortuosities $\tau_0$ and $\tau_{\text{end}}$. The values of $\tau_{\text{end}}$ are depicted with the colored bars and related to the left ordinate axis. The values of $\tau_0$ are depicted with the hatched bars and related to the right ordinate axis. The meaning of the colors and line types is similar to those from \autoref{fig:Duration}. }
	\label{fig:Tortuosity}
\end{figure}

The influence on the geodesic tortuosities as a measure for the effective conductivity is shown in \autoref{fig:Tortuosity}. Adding binder in general increases the tortuosity by approximately 10\,\%.  Moreover,  $\tau_{\text{end}}$ behaves inversely proportional to $S^{\mathrm{E}}_{\text{final}}$ (cf.\ \autoref{fig:Saturation}). Thus, the transport pathways elongate when more gas agglomerates are entrapped. For most electrodes with $S^{\mathrm{E}}_{\text{final}} > 90\,\%$ the influence is minor. However, in the extreme case (ID~14) the shortest pathway increases by 27.7\,\%. 

\begin{figure}
  \centering
      \includegraphics[width=0.6\textwidth]{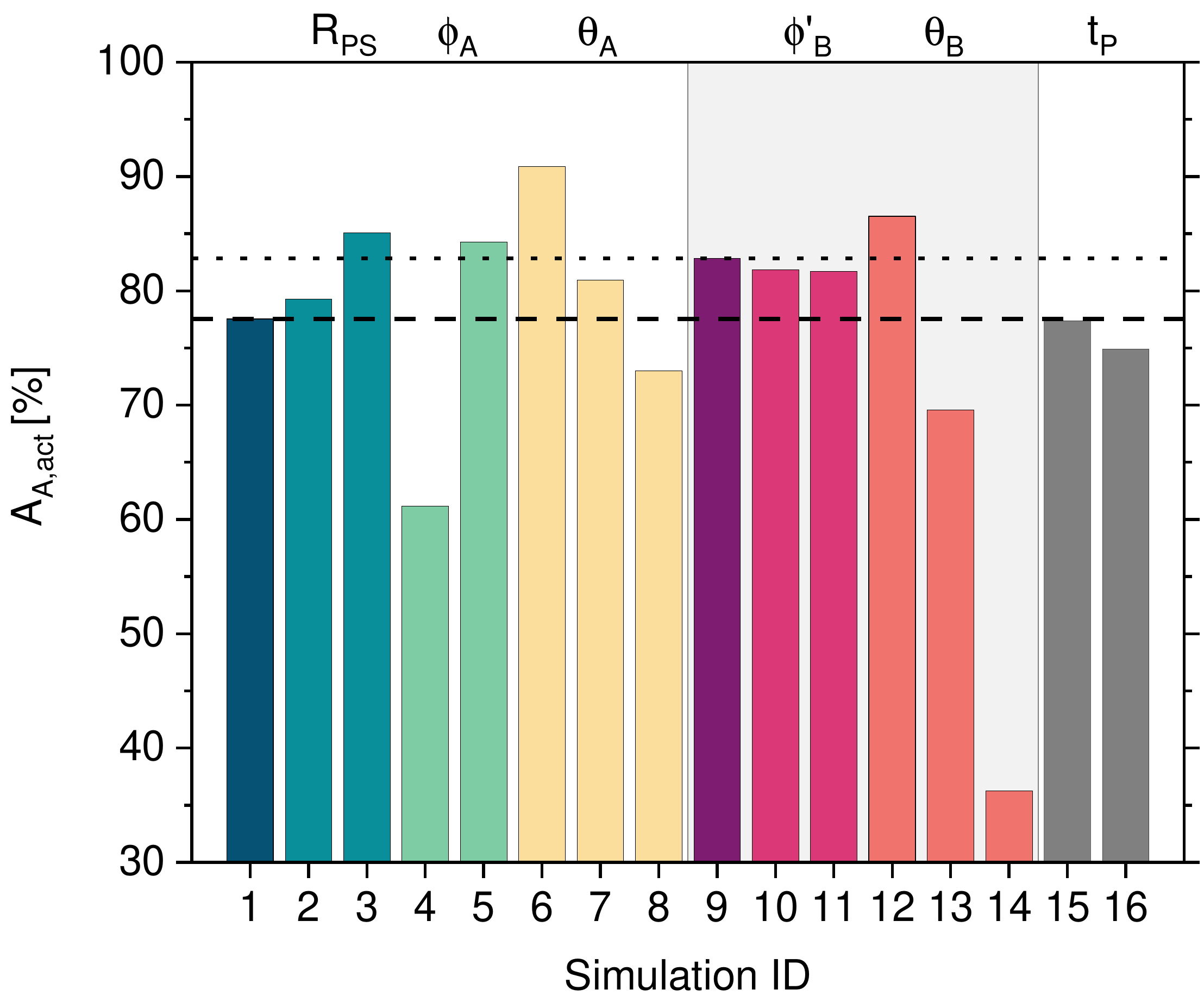}
    \caption{Overview of the ratio of the electrochemically active surface area $A_{\text{A,active}}$. The meaning of the colors is identical to those from \autoref{fig:Duration}. The horizontal dashed and dotted line represent $A_{\text{A,act}}$ of ID~1 and ID~9, respectively. }
  \label{fig:GasArea}
\end{figure}

The influence of entrapped gas on the electrochemically active surface area $A_{\text{A,act}}$ is shown in \autoref{fig:GasArea}. As expected a direct proportionality of $A_{\text{A,act}}$ and $S^{\mathrm{E}}_{\text{final}}$ is observed. Better saturation and less residual gas phase decrease the surface area of active material being in contact with electrolyte. However, the quantities are surprising. Even for the best saturation (ID~6) and a strongly hydrophilic, i.e.\ gas repelling, surface, about 9\,\% of the total active surface area are passivated. For the worst case (ID~14) even 63.8\,\% of active surface area are blocked.

Note that these results represent gas entrapment right after filling. It might differ from the gas entrapment at the end of the whole manufacturing process during which gas is either removed by evacuation or in subsequent production steps.

The aforementioned results (cf.\ \figsref{fig:Tortuosity}{fig:GasArea}) confirm that electrode design and filling process have a huge effect on battery performance. As was shown previously in experiments \cite{Davoodabadi2020,Sauter2020} and simulations \cite{Sauter2020,Shodiev2021}, especially structural properties of the electrodes play an important role. The larger the pores are and the better they are connected, the better is the effective ionic conductivity and the more surface area remains electrochemically active. These effects can even be enhanced when increasing the wettability of electrode components. Thus, the results indicate, that increasing the power density by calendering electrodes increases the amount of entrapped gas which leads to a tortuosity increase and finally reduces battery capacity.

\subsection{Permeability} \label{subsec:PermResults} 
\begin{figure}
	\centering
	  \includegraphics[width=0.7\textwidth]{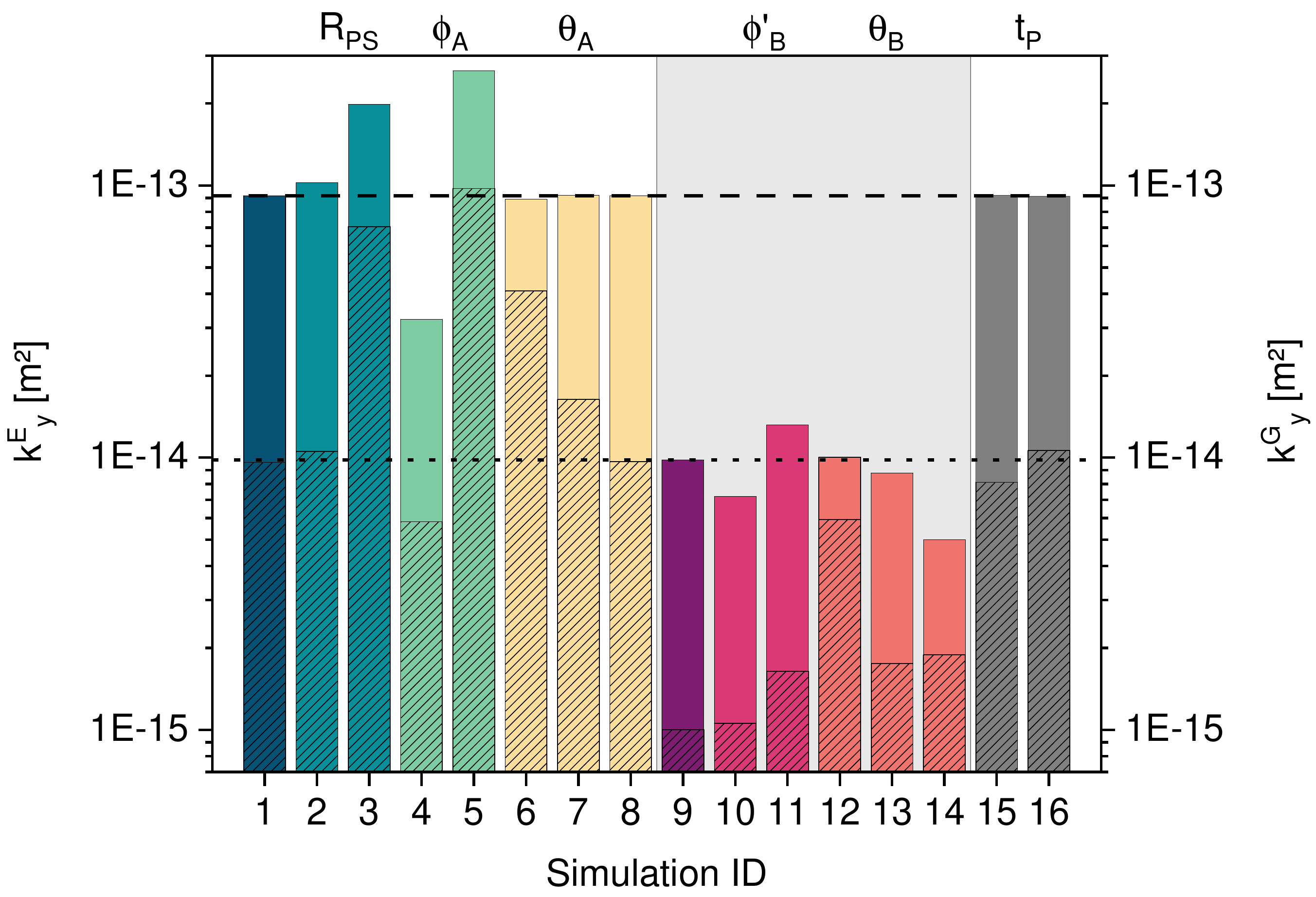}
	\caption{Overview of the electrolyte permeability $k^{\mathrm{E}}_y$ and the gas permeability $k^{\mathrm{G}}_y$ at the end of the filling process. The values of $k^{\mathrm{E}}_y$ are depicted with the colored bars and related to the left ordinate axis. The values of $k^{\mathrm{G}}_y$ are depicted with the hatched bars and related to the right ordinate axis. The meaning of the colors is identical to those from \autoref{fig:Duration}. The horizontal dashed and dotted line represent $k^{\mathrm{E}}_y$ of ID~1 and ID~9, respectively. }
	\label{fig:Permeability}
\end{figure}

The permeabilities of electrolyte and residual gas phase at the end of the filling process are shown in \autoref{fig:Permeability}. Numerical results are given in \autoref{tab:SimulationOverview}.

The general observations from \autoref{fig:Permeability} are: 1)~Permeabilities for electrodes without binder (IDs~1$-$8, 15, and 16) are about one order of magnitude larger than for cases with binder (IDs~9$-$14). 2)~Gas permeabilities are mostly about one order of magnitude smaller than electrolyte permeabilities.

Both effects are mainly influenced by the solid-fluid interfacial contact area and the connectivity of the fluid phase \cite{Landry2014}. An increasing solid-fluid contact area increases the flow resistance, thus reducing the fluid mobility \cite{Li2005,Landry2014}. In contrast, a better connectivity enhances the mobility \cite{Avraam1995,Landry2014}. Electrodes without binder, and thus with less solid material, exhibit a smaller specific solid-fluid contact area, and lead to larger permeabilities. The residual gas phase, which has a remarkably smaller volume fraction than the electrolyte, has a low connectivity, and thus a lower permeability.

Beside the two aforementioned parameters, also structural properties \cite{Dou2013,Zhang2016,Ahkami2020}, fluid saturation, wettability of the solids \cite{Landry2014,Li2005,Dou2013,Ghassemi2011}, and fluid-fluid interfacial area \cite{Avraam1995,Landry2014,Li2019} affect the permeability. Apart from the fluid-fluid interfacial area, all other effects are shown in \autoref{fig:Permeability} and are discussed in the following.

Large $R_{\text{PS}}$ (ID~3) or small $\phi_{\text{A}}$ (ID~5) result in large values of $k^{\mathrm{E}}_y$ and $k^{\mathrm{G}}_y$. In both cases the pores are comparable in size which leads to a small solid-fluid contact area and a low flow resistance \cite{Ahkami2020}. Moreover, the amount of residual gas phase and its connectivity is low for both structures (cf.\ \autoref{fig:Saturation}). Although, this typically decreases the permeability, here, the effect is dominated by drag of the electrolyte phase leading to large values of $k^{\mathrm{G}}_y$, too. This has already been observed experimentally \cite{Avraam1995}. It is also reproduced by the electrode structures with binder (IDs~9$-$11) where the permeability increases for larger inner volume fractions.

An increasing wettability results from strong solid-fluid adhesion forces. Thus, the wetting phase is highly attracted by the solid, increases the solid-fluid interface, and decreases fluid mobility. This is different for the nonwetting phase \cite{Landry2014,Lautenschlaeger2020}. However, wettability and fluid connectivity are competing effects \cite{Landry2014}. This is also shown in \autoref{fig:Permeability} where the electrolyte permeability is hardly affected by the wettability of active material (IDs~1, 6$-$8). But when also taking into account the electrolyte saturation $S^{\mathrm{E}}_{\text{final}}$ (cf.\ \autoref{fig:Saturation}) the link is not so clear anymore. As the contact angle $\theta_{\text{A}}$ increases, the saturation decreases and thereby reduces the electrolyte connectivity. Thus, an apparent effect is observed here. In fact, the enhanced electrolyte mobility for increasing contact angles is compensated by a decreased connectivity. This effect is even more pronounced for simulations in which the binder wettability was varied (IDs~9, 12$-$14). There, increasing $\theta_{\text{B}}$ counter-intuitively reduces the electrolyte permeability. The same two competing effects affect the gas phase. Thus, leading to an increased gas permeability for a better electrolyte wettability.

\section{Conclusion}  \label{sec:Conclusion}
In this study, LBM simulations were used to improve the understanding of electrolyte filling processes on the pore scale. Therefore, a new lattice Boltzmann model for studying multi-phase fluid flow simultaneously in pores of different length scales is presented. This model was applied to study  electrolyte filling of realistic 3D lithium-ion cathodes with and without binder. Similarly other battery components as well as complete cells can be analyzed. The methodology is universal and can also be applied to other energy storage devices such as metal-air batteries, flow batteries or fuel cells. The influence of a wide range of relevant structural and physico-chemical properties as well as process parameters was studied. Large-scale simulations were conducted in which the particle size, volume fraction, and wettability of active material, the distribution, inner volume fraction, and wettability of binder, as well as the process time were varied. Pressure-saturation curves were determined. They show a systematic entrapment of residual gas that depends on the aforementioned parameters. A detailed analysis was conducted to understand the interdependencies of the amount, spatial distribution, and size of the gas agglomerates, as well as their effect on transport properties and electrochemically active surfaces.

In general, the findings indicate that the filling process is mainly influenced by structural electrode properties. It can be optimized by increasing the wettability. The influence of the process time is subordinate for the values studied here. At the end of the filling process, most electrodes contained 10\,\% or less residual gas phase. It was shown that large pores with a narrow pore size distribution and hydrophilic active material reduce gas entrapment. It could be further reduced when adding a wetting binder. Increasing the filling speed resulted in the entrapment of a slightly larger amount of medium-sized gas agglomerates. The worst saturation was observed for adding a dewetting binder.

A detailed analysis of the position and distribution of gas agglomerates was conducted and correlated with the battery performance. It was shown that gas agglomerates increase ionic transport pathways in electrodes and thus reduce the effective ionic conductivity. Moreover, gas agglomerates decrease the electrochemically active surface area. Both effects increase overpotentials during battery operation and have a negative impact on the specific battery capacity. The most favorable results were observed for electrodes with large pores, good pore space connectivity, and good wettability of electrode components. The results indicate that calendering electrodes could potentially reduce the power density of batteries.

Finally, it was shown which efforts are necessary to displace gas agglomerates from electrodes. For that, electrolyte and gas permeabilities at the end of the filling process were determined. The findings indicate that the binder decreases the mobility of gas agglomerates. The largest permeabilities were observed for large pores with a narrow pore size distribution and a wetting active material.

Altogether, it is shown that the new lattice Boltzmann model yields a detailed insight and a profound understanding of the influencing factors of filling processes on the pore scale. The results are promising and can especially be used to support electrode and electrolyte design as well as for optimizing the filling process. 

\section*{Acknowledgement}  \label{sec:Acknowledgement}
The authors gratefully acknowledge financial support from the European Union's Horizon 2020 Research and Innovation Programme within the project ``DEFACTO'' [grant number 875247]. Furthermore, the presented work was financially supported by the Bundesministerium für Bildung und Forschung (BMBF) within the project HiStructures [grant number 03XP0243D]. The simulations were carried out on the Hawk at the High Performance Computing Center Stuttgart (HLRS) [grant LaBoRESys], and on JUSTUS 2 at the University Ulm [grant INST 40/467-1 FUGG].

\clearpage
\begin{appendices}
\renewcommand{\appendixautorefname}{Section}
\renewcommand{\theequation}{(\thesection.\arabic{equation})}
\setcounter{equation}{0}
\section{LBM Details} \label{sec:Appendix_LBM}

\subsection{D3Q19 Velocity Set}
The D3Q19 velocity set used in the present work is
\begin{align}
    \label{eq:VelocitySet}
    &{\left[\mathbf{c}_{0}, \mathbf{c}_{1}, \mathbf{c}_{2}, \mathbf{c}_{3}, \mathbf{c}_{4}, \mathbf{c}_{5}, \mathbf{c}_{6}, \mathbf{c}_{7}, \mathbf{c}_{8}, \mathbf{c}_{9}, \mathbf{c}_{10}, \mathbf{c}_{11}, \mathbf{c}_{12}, \mathbf{c}_{13}, \mathbf{c}_{14}, \mathbf{c}_{15} \mathbf{c}_{16}, \mathbf{c}_{17}, \mathbf{c}_{18}\right]} \\
    &\quad=\frac{\Delta x}{\Delta t}\left[\begin{array}{rrrrrrrrrrrrrrrrrrr}
        0 & -1 &  0 &  0 & 1 & 0 & 0 & -1 & -1 & -1 & -1 &  0 &  0 & 1 &  1 & 1 &  1 & 0 &  0 \\
        0 &  0 & -1 &  0 & 0 & 1 & 0 & -1 &  1 &  0 &  0 & -1 & -1 & 1 & -1 & 0 &  0 & 1 &  1 \\
        0 &  0 &  0 & -1 & 0 & 0 & 1 &  0 &  0 & -1 &  1 & -1 &  1 & 0 &  0 & 1 & -1 & 1 & -1 
    \end{array}\right]
   \nonumber
\end{align}

\subsection{Equilibrium Distribution Function}
The Maxwell-Boltzmann equilibrium distribution function is
\begin{align}
    \label{eq:Maxwell}
    f^{eq}_i\left(\rho,\mathbf{u}\right) = w_i \rho \left[1 + \frac{\mathbf{c}_i \mathbf{u}}{c_s^2} + \frac{(\mathbf{c}_i \mathbf{u})^2}{2 c_s^4} - \frac{\mathbf{u} \mathbf{u}}{2 c_s^2}\right].
\end{align}
Here, $w_i$ are the lattice specific weights ($w_i=1/3$ for $i=0$, $w_i=1/18$ for $i=1...6$, and $w_i=1/36$ for $i=7...18$). The lattice speed of sound is $c_{s}=1/\sqrt{3}$. 

\subsection{Physical Quantities}
From the distribution function $\boldsymbol{f}$, different relevant macroscopic properties can be determined locally. Those are, e.g., the density 
\begin{align}
    \label{eq:rho_BGK}
    \rho = \sum_{i}^{}f_i,
\end{align}
and the macroscopic velocity
\begin{align}
    \label{eq:mom_BGK}
    \textbf{u} = \frac{1}{\rho}\sum_{i}^{}f_i\textbf{c}_i.
\end{align}

The total pressure of the mixture follows the ideal gas law and is extended by the interaction contributions of both components:
\begin{align}    
    \label{eq:SC_pressure}
    p(\mathbf{x}) = c_{s}^2 \left[\rho(\mathbf{x}) + G^{\sigma\bar{\sigma}}_{\mathrm{inter}}\rho^{\sigma}(\mathbf{x})\rho^{\bar{\sigma}}(\mathbf{x}) \Delta t^2\right].
\end{align}

\subsection{Correlation for the Contact Angle}
Following Huang~\textit{et al.}\ \cite{Huang2007}, the interaction parameters of the wetting (w) and nonwetting (nw) phase are typically related as $G^{\mathrm{nw}}_{\mathrm{ads}} = - G^{\mathrm{w}}_{\mathrm{ads}}$ leading to the the contact angle

\begin{align}
    \label{eq:contactAngle}
   \cos\theta = \frac{4G^{\mathrm{nw}}_{\mathrm{ads}}}{G_{\mathrm{inter}}(\rho^{\mathrm{w}}-\rho_{\mathrm{dis}}^{\mathrm{nw}})}.
\end{align}

\end{appendices}
\unappendix
\clearpage

\bibliographystyle{elsarticle-num} 

\bibliography{main}


\pagebreak

\begin{center}
  \begin{singlespace}
  \Large Supporting Information: Understanding Electrolyte Filling of Lithium-Ion Battery Electrodes on the Pore Scale Using the Lattice Boltzmann Method\\[.5cm]
  \normalsize Martin P. Lautenschlaeger$^{a,b,*}$, Benedikt Prifling$^{c}$, Benjamin Kellers$^{a,b}$, Julius Weinmiller$^{a,b}$, Timo Danner$^{a,b}$, Volker Schmidt$^{c}$, Arnulf Latz$^{a,b,d}$\\[.3cm]
  {\itshape \small ${}^a$German Aerospace Center (DLR), Institute of Engineering Thermodynamics, 70569 Stuttgart, Germany\\
  ${}^b$Helmholtz Institute Ulm for Electrochemical Energy Storage (HIU), 89081 Ulm, Germany\\
  ${}^c$Ulm University (UUlm), Institute of Stochastics, 89081 Ulm, Germany\\
  ${}^d$Ulm University (UUlm), Institute of Electrochemistry, 89081 Ulm, Germany\\[.3cm]}
  \small ${}^*$Corresponding author: Martin.Lautenschlaeger@dlr.de\\[1cm]
  \end{singlespace}
\end{center}

\setcounter{section}{0}
\setcounter{equation}{0}
\setcounter{figure}{0}
\setcounter{table}{0}
\setcounter{page}{1}
\renewcommand{\thesection}{S\arabic{section}}
\renewcommand{\theequation}{(S\arabic{equation})}
\renewcommand{\thefigure}{S\arabic{figure}}
\renewcommand{\thetable}{S\arabic{table}}

In the present paper, electrolyte filling processes were studied by means of the lattice Boltzmann method (LBM) with regard to the influence of structural and physico-chemical properties as well as the process time $t_{\mathrm{P}}$. In particular, the influencing factors were the particle size distribution $R_{\mathrm{PS}}$, the volume fraction $\phi_{\mathrm{A}}$, and the wettability $\theta_{\mathrm{A}}$ of the active material as well as the the inner volume fraction $\phi'_{\mathrm{B}}$ and the wettability $\theta_{\mathrm{B}}$ of the binder. Results were reported as pressure-saturation relationship, final degree of saturation, detailed analysis of the gas entrapment, and permeability. The numerical values of the results as well as complementary information are given in the following.

The Supporting Information is organized as follows. In \autoref{SI-sec:LBM} the general LBM and the multi-component Shan-Chen pseudopotential method (MCSC) are briefly described. A verification of the combined MCSC and grayscale (GS) model is given in \autoref{SI-sec:ModelVerification}. The uncertainty estimation of the pressure-saturation curves is described in \autoref{SI-sec:UncertPressSat}. The numerical values of the results presented in the present paper are summarized in \autoref{SI-sec:NumericalResults}. Finally, additional plots of the size distributions of gas agglomerates for all simulations from the present paper are shown in \autoref{SI-sec:GasEntrapment}.

\clearpage

\section{LBM} \label{SI-sec:LBM} 
\subsection{General introduction} \label{SI-subsec:LBM} 
The book \textit{The Lattice Boltzmann Method} \cite{Krueger2016} introduces LBM in very detail. It is also helpful to get a comprehensive overview over the method and its applications. In the following, only a condensed overview of the lattice Boltzmann models that are relevant for the present paper is given. 

The general LBM for single-phase fluid flow solves the discretized Boltzmann equation  
\begin{align}
    \label{SI-eq:LBM}
    \frac{\partial f_i\left(\mathbf{x},t\right)}{\partial t} + \mathbf{c}_i \nabla f_i\left(\mathbf{x},t\right) = \Omega_i\left(\mathbf{x},t\right),
\end{align}
where $\boldsymbol{f}$ are the distribution functions, $\Omega$ is a general collision operator, and $\boldsymbol{x}$ and $t$ denote the position of the lattice cell and the time, respectively. As already described in the main text and the appendix of the present paper, \autoref{SI-eq:LBM} is discretized on a regular and cubic 3D lattice using the D3Q19 velocity set. The directions of the velocity set are denoted as $i$. They correspond to the directions along which information from the distribution functions $\boldsymbol{f}$ is transferred. They are directly connected to the predefined lattice velocity $\boldsymbol{c}_i$, i.e.\ the microscopic speed of transport during a single time step $\Delta t$. The collision operator $\Omega$ describes the physics of the problem via particle collisions which lead to modifications and a redistribution of $\boldsymbol{f}$. The simplest and most commonly used functional form of $\Omega$ is from Bhatnagar, Gross, and Krook (BGK) \cite{Bhatnagar1954}
\begin{align}
    \label{SI-eq:BGK}
    \Omega_i = - \frac{1}{\tilde{\tau}} \left(f_i-f^{\mathrm{eq}}_i \right).
\end{align}

The combination of \eqsref{SI-eq:LBM}{SI-eq:BGK} is referred to as the lattice BGK (LBGK) equation. It describes the relaxation of $\boldsymbol{f}$ towards the Maxwell–Boltzmann equilibrium distribution function $\boldsymbol{f}^{\mathrm{eq}}$ (cf.\ appendix of the present paper). The characteristic relaxation time is denoted by $\tilde{\tau}$.

By solving \autoref{SI-eq:LBM}, different relevant macroscopic properties can be determined locally as moments of $\boldsymbol{f}$. Examples are given in the appendix of the present paper.

In addition, it is also important to model interaction between the fluid and the solid wall since fluid flow through porous electrode structures is studied in the present paper. The simplest and most popular approach for such a no-slip boundary condition is the bounce-back method \cite{Chen1998,Liu2016}. Using this method, distribution functions that approach the wall are reflected back to the lattice cell from which they originated. There are different types of bounce-back schemes reported in the LBM literature, where the so-called halfway bounce-back scheme is applied here \cite{Krueger2016}. This frequently used approach is defined as
\begin{align}
    \label{SI-eq:bounceback}
    f_i(\mathbf{x},t) = f_{\bar{i}}(\mathbf{x},t+\Delta t),
\end{align}
where $\bar{i}$ denotes the direction opposite to $i$, i.e.\ $\boldsymbol{c}_{\bar{i}}=-\boldsymbol{c}_i$.

\subsection{Multi-Component Shan-Chen Pseudopotential Method} \label{SI-subsec:MCSC} 
The MCSC \cite{Shan1993} can be applied to study multi-phase fluid flows. It is based on a bottom-up modeling approach \cite{Chen2014,Liu2016} in which molecular interaction forces are determined from the pseudopotential $\psi = f(\rho)$. In the following, the model is described for two immiscible components and the pseudopotential $\psi = \rho$, which is a typical choice in the literature \cite{Pan2004,Huang2007,Li2018,Jeon2019,Pereira2019,Shodiev2021}.

In the MCSC, each lattice cell is occupied by all immiscible components. The temporal evolution of $\boldsymbol{f}^{\sigma}$ is described by the lattice Boltzmann (LB) equation (cf.\ \autoref{SI-eq:LBM}) with the BGK collision operator (cf.\ \autoref{SI-eq:BGK}). Using the Shan-Chen forcing approach, one obtains
\begin{align}
    \label{SI-eq:SC_BGK}
    f^{\sigma}_i\left(\mathbf{x}+\mathbf{c}_i \Delta t, t + \Delta t \right) - f^{\sigma}_i\left(\mathbf{x}, t\right) = - \frac{\Delta t}{\tilde{\tau}^{\sigma}} \left(f^{\sigma}_i\left(\mathbf{x}, t\right) - f^{\mathrm{eq},\sigma}_i\left(\mathbf{x}, t\right)\right),
\end{align}
where $\sigma$ denotes the component, i.e.\ electrolyte or gas phase. 

In addition, the interfacial tension between the components $\sigma$ and $\bar{\sigma}$ is modeled as a fluid-fluid interaction force $\boldsymbol{F}^{\sigma}_\mathrm{inter}$. The wettability or adhesion at a solid wall is modeled as a solid-fluid interaction force $\boldsymbol{F}^{\sigma}_\mathrm{ads}$. The external force fields $\boldsymbol{F}^{\sigma}_\mathrm{ext}$ can be considered in the simulations, where the definition of all three forces is given in the main text. The sum of the aforementioned force contributions determines the total force $\boldsymbol{F}^{\sigma}_\mathrm{tot}=\boldsymbol{F}^{\sigma}_\mathrm{inter}+\boldsymbol{F}^{\sigma}_\mathrm{ads}+\boldsymbol{F}^{\sigma}_\mathrm{ext}$ acting on a lattice cell. Using the Shan-Chen forcing approach \cite{Krueger2016}, $\boldsymbol{F}^{\sigma}_\mathrm{tot}$ is finally incorporated into MCSC as a force-induced contribution to the equilibrium velocity of each component. More precisely, it holds
\begin{align}    
    \label{SI-eq:SC_u_eq}
    \textbf{u}^{\mathrm{eq},\sigma} = \frac{\sum_{\sigma}^{}\rho^{\sigma}\mathbf{u}^{\sigma}/\tilde{\tau}^{\sigma}}{\sum_{\sigma}^{}\rho^{\sigma}/\tilde{\tau}^{\sigma}} + \frac{\tilde{\tau}^{\sigma}\boldsymbol{F}^{\sigma}_\mathrm{tot}}{\rho^{\sigma}}.
\end{align}

The equilibrium velocity $\boldsymbol{u}^{\mathrm{eq},\sigma}$ must not be confused with the macroscopic streaming velocity of the mixture. The latter has also to be force-corrected and is given by
\begin{align}    
    \label{SI-eq:SC_u_macro}
    \textbf{u}_\mathrm{macro} = \sum_{\sigma}^{}\left(\sum_{i}^{} \frac{f^{\sigma}_i \mathbf{c}_i}{\rho^{\sigma}} + \frac{\boldsymbol{F}^{\sigma}_\mathrm{tot} \Delta t}{2 \rho^{\sigma}} \right).
\end{align}

\section{Verification of Model Parameters} \label{SI-sec:ModelVerification}

The LB model that was applied for the current study is described in Section 2.1 in the present paper. Compared to the model proposed by Pereira \cite{Pereira2016,Pereira2017,Pereira2019} it contains some adaptions and uses a different forcing scheme. Therefore, potential effects of the model changes on the physical behavior of the model were tested. Verifications with respect to the interfacial tension and the wetting behavior were conducted. They are described in the following.

\subsection{Interfacial Tension} \label{SI-subsec:InterfacialTension}

Fluid flow through a homogenized binder region should not affect the interfacial tension between the electrolyte and the gas phase. Thus, it has to be ensured that setting the same value for $G^{\mathrm{EG}}_\mathrm{inter}$ in all lattice cells of the system does not lead to different interfacial tensions.

\begin{figure}
	\centering
	  \includegraphics[width=0.5\textwidth]{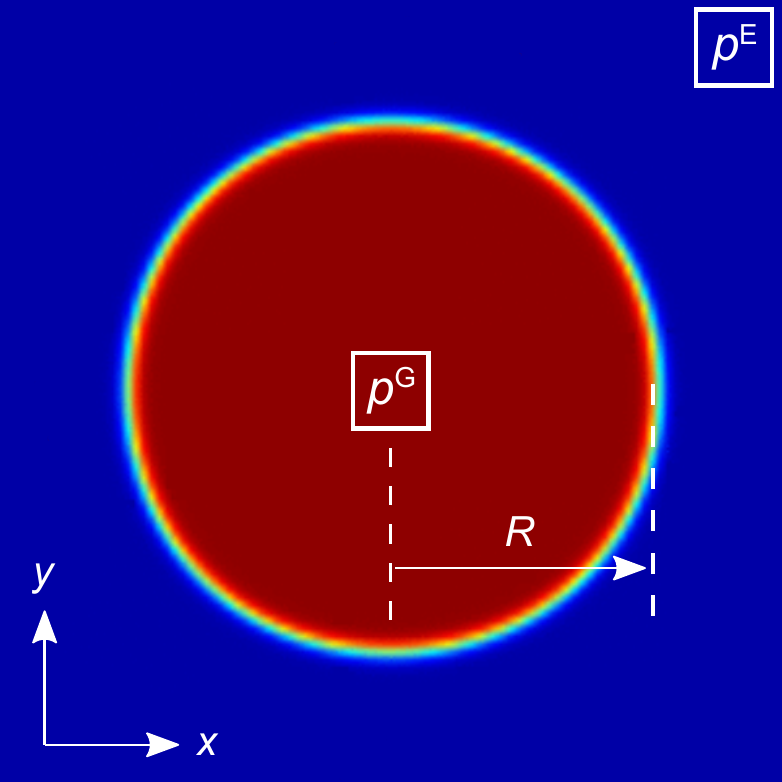}
	\caption{Simulation setup for evaluating the influence of the homogenized model on the interfacial tension. A gas bubble (red) with radius $R$ is submersed in an electrolyte phase (blue) of equal mass. The pressures of both the gas phase $p^\mathrm{G}$ and the electrolyte $p^\mathrm{E}$ are determined at the locations indicated by the white boxes.}
	\label{SI-fig:Laplace}
\end{figure}

Therefore, a series of bubble tests was conducted from which the Laplace pressure was determined. The simulation setup is shown in \autoref{SI-fig:Laplace} and consists of a fully periodic 2D system with a size of 100 cells along the $x$- and $y$-direction. The system contains a gas bubble with the density $\rho^{\mathrm{G}}$ which is surrounded by electrolyte with the density $\rho^{\mathrm{E}}$. Both components have equal masses. The model parameters were similar to those given in Table 1 in the present paper. Here, in each cell, the homogenized model, i.e.\ the combined MCSC and GS method, was applied. The values of the solid-fluid interaction parameter $G_\mathrm{ads}=G^{\mathrm{G}}_\mathrm{ads}=-G^{\mathrm{E}}_\mathrm{ads}$ and the solid fraction $n_s$ were were identical in each cell. They were varied between the simulations.

A simulation run consisted of 500,000 time steps in which the pressure difference between the gas bubble and the electrolyte $\Delta p=p^\mathrm{G}-p^\mathrm{E}$ as well as the bubble radius $R$ were determined every 10,000 time steps. Therefrom, the interfacial tension $\gamma$ was determined using Laplace's law $\gamma = \Delta p R$. 

The results are given in \autoref{SI-tab:BubbleTest}. It can be shown that there is no influence of $G_{\mathrm{ads}}$ and hardly any impact of $n_s$ on $\gamma$. The values of $\gamma$ are in good agreement with the value of $\gamma$ used in this study (cf.\ Table 1 in the present paper). Therefrom, it is concluded that the model adaptions for homogenized components, i.e.\ $n_s\neq0$, do not lead to different interfacial tensions compared to the pure MCSC (cf.\ \cite{Huang2007}). Another advantage of the present model is that the scaling of the interfacial force parameter $G_\mathrm{inter}$ also ensures a stable and non-diverging interface (cf.\ \autoref{SI-fig:Laplace}).

\begin{table}[h!]
    \centering
    \caption{Results of the surface tensions $\gamma$ determined from the bubble test simulations. The parameter $n_s$ is the solid fraction of the homogenized lattice cell, $G_{\mathrm{ads}}$ is the solid-fluid interaction parameter, $R$ is the bubble radius, and $\Delta p$ is the Laplace pressure. }
    \begin{tabular}{c|c|c|c|c}
    \toprule \toprule
        $n_s$ & $G_{\mathrm{ads}}$ & $R$ (m) & $\Delta p$ (Pa) & $\gamma$ (N/m) \\ 
        \midrule
        0.0 & 0.0 & 1.487E-05 & 2782.64 & 0.04137 \\ 
        0.4 & 0.0 & 1.487E-05 & 2847.24 & 0.04234 \\ 
        0.4 & 0.2 & 1.487E-05 & 2847.24 & 0.04234 \\ 
        0.4 & 0.4 & 1.487E-05 & 2847.24 & 0.04234 \\ 
        0.5 & 0.0 & 1.488E-05 & 2842.64 & 0.04230 \\ 
        0.5 & 0.2 & 1.488E-05 & 2842.64 & 0.04230 \\ 
        0.5 & 0.4 & 1.488E-05 & 2842.64 & 0.04230 \\ 
        0.6 & 0.0 & 1.488E-05 & 2838.23 & 0.04222 \\ 
        0.6 & 0.2 & 1.488E-05 & 2838.23 & 0.04222 \\ 
        0.6 & 0.4 & 1.488E-05 & 2838.23 & 0.04222 \\ 
        0.9 & 0.0 & 1.488E-05 & 2804.23 & 0.04174 \\ 
    \bottomrule \bottomrule
    \end{tabular}
    \label{SI-tab:BubbleTest}%
\end{table}

\subsection{Wetting Behavior} \label{SI-subsec:WettingBehavior}

The inner volume fraction of the binder $\phi'_{\text{B}} = n_s$ should not affect the contact angle $\theta$ between the binder and the electrolyte or the gas, respectively. Thus, it has to be ensured that the choice of the solid-fluid interaction parameter $G_{\mathrm{ads}}=G^{\mathrm{G}}_\mathrm{ads}=-G^{\mathrm{E}}_\mathrm{ads}$ which determines $\theta$, is independent of $n_s$.

Typically, the correlation between the solid-fluid interaction force $G_\mathrm{ads}$ and the contact angle $\theta$ is determined via contact angle measurements on solid surfaces. This is not possible when using solids with inner porosity, where the fluid is either fully repelled or fully absorbed into the solid. Therefore, the effect of the homogenized model on the contact angle was studied using an approach similar to the Washburn experiment \cite{Washburn1921}. A tube or channel is filled in equal parts with electrolyte and gas, where the electrolyte is absorbed into the channel by attractive solid-fluid interaction forces. The speed of the absorption, i.e.\ the velocity of the advancing electrolyte-gas interface $v_x=\mathrm{d}\mathbf{x}/\mathrm{d}t$ correlates with $G_\mathrm{ads}$. 

This correlation can be derived from Hagen-Poiseuille's law for a capillary with radius $R_{\mathrm{pore}}$ and the capillary pressure $\Delta p=2\gamma\cos(\theta)/R_{\mathrm{pore}}$:
\begin{align}
    \label{SI-eq:HagenPoiseuille}
    \begin{split}
    \frac{\mathrm{d}V}{\mathrm{d}t} = \frac{\mathrm{d}\mathbf{x} (\pi R^2_{\mathrm{pore}})}{\mathrm{d}t} &= \frac{\pi R^4_{\mathrm{pore}} \Delta p}{8 \eta L}\\\
    v_x = \frac{\mathrm{d}\mathbf{x}}{\mathrm{d}t} &= \frac{R_{\mathrm{pore}} \gamma\cos(\theta)}{4 \eta L}.
    \end{split}
\end{align}
Here, $\mathrm{d}V/\mathrm{d}t$ is the volumetric flow rate, $\eta$ is the dynamic viscosity of the electrolyte, and $L$ is the length of the capillary.

The simulation setup consists of a 2D system with a size of $L=500$ cells along the $x$- and $H=5$ cells along the $y$-direction. Each cell was handled the same way as the binder in the present paper, i.e.\ the homogenized model was applied. A constant density was prescribed at boundaries in $x$-direction, whereas periodic boundary conditions were applied in $y$-direction. The left and the right half of the system were initialized with an electrolyte with density $\rho^{\mathrm{E}}$ and a gas phase with density $\rho^{\mathrm{G}}$, respectively. Both components had equal masses. No pressure gradient or other external force field was applied. The model parameters were similar to those given in Table 1 in the present paper. 

A simulation run consisted of 1,500,000 time steps. Due to the attractive adhesive forces on the electrolyte, the interface started moving along the $+x$-direction, i.e.\ electrolyte displacing gas. The interface velocity $v$ was determined every 10,000 time steps. 

The solid fraction $n_s$ was varied in the range $n_s = [0.3,\,0.7]$. This range includes the values of $n_s$ that were chosen in the present paper. For each value of $n_s$, a broad range of values for $G_{\mathrm{ads}}=G^{\mathrm{G}}_\mathrm{ads}=-G^{\mathrm{E}}_\mathrm{ads}$ was considered and the velocity of the advancing electrolyte-gas interface $v$ was recorded. The contact angle $\theta$ was then determined by inserting $v$ into \autoref{SI-eq:HagenPoiseuille}. The corresponding results of $\theta$ are given in \autoref{SI-tab:Washburn}. They indicate that the contact angle mainly depends on $G_{\mathrm{ads}}$ and is hardly affected by $n_s$. However, $n_s$ correlates with the capillary radius $R_{\mathrm{pore}}$ in the range between $R=132$\,nm and $R=290$\,nm which agrees with experimental values from the literature \cite{Daemi2018}.

\begin{table}[h!]
    \centering
    \caption{Results of the contact angle $\theta$ determined from the capillary simulations. The parameter $n_s$ is the solid fraction of the homogenized lattice cell, $R_{\mathrm{pore}}$ is the capillary radius, and $G_{\mathrm{ads}}$ is the solid-fluid interaction parameter.  }
    \begin{tabular}{c|c||l c|c|c|c|c|c|c|c}
    \toprule \toprule
        $n_s$ & $R_{\mathrm{pore}}$ (nm) & $G_{\mathrm{ads}}=$ & 0.05 & 0.10 & 0.15 & 0.175 & 0.20 & 0.25 & 0.30 & 0.32 \\ 
        \midrule
        0.3 & 290 & $\theta~(\degree)=$ & 81.96 & 73.36 & 64.28 & 63.66 & 54.41 & 45.05 & 29.14 & 19.57 \\ 
        0.4 & 248 &  & 81.75 & 73.12 & 64.18 & 59.49 & 54.48 & 42.85 & 29.04 & 22.15 \\ 
        0.5 & 220 &  & 82.06 & 73.80 & 65.17 & 60.64 & 55.68 & 44.32 & 29.27 & 20.39 \\ 
        0.6 & 180 &  & 82.11 & 73.96 & 65.37 & 60.80 & 55.92 & 44.79 & 29.80 & 20.97 \\ 
        0.7 & 132 &  & 82.11 & 73.96 & 65.37 & 60.80 & 55.92 & 44.79 & 29.80 & 20.97 \\ 
    \bottomrule \bottomrule
	\end{tabular}%
	\label{SI-tab:Washburn}%
\end{table}

The values of $G_{\mathrm{ads}}$ that were chosen for this study were taken from \autoref{SI-tab:Washburn}. In particular, we used $G_{\mathrm{ads}} = \{0.3,\,0.175,\,0.0,\,-0.175\}$ for the corresponding binder contact angles $\theta_{\text{B}} = \{30,\,60,\,90,\,120\}\degree$.

\section{Uncertainty Estimation for Pressure-Saturation Behavior} \label{SI-sec:UncertPressSat} 

Filling simulations were conducted for five different, statistically equivalent electrode realizations which had the same macroscopic electrode properties. All electrode realizations correspond to a medium particle size distribution $R_{\mathrm{PS}} = \mathrm{medium}$, the volume fraction of the active material $\phi_{\mathrm{A}}=0.6$, wettability $\theta_{\mathrm{A}}=90\degree$, and no binder content. Separate pressure-saturation curves were determined from each simulation and for filling along different directions (to study also the effect of structural anisotropy) and an average pressure-saturation curve was derived.

\begin{figure}
	\centering
	  \includegraphics[width=0.8\textwidth]{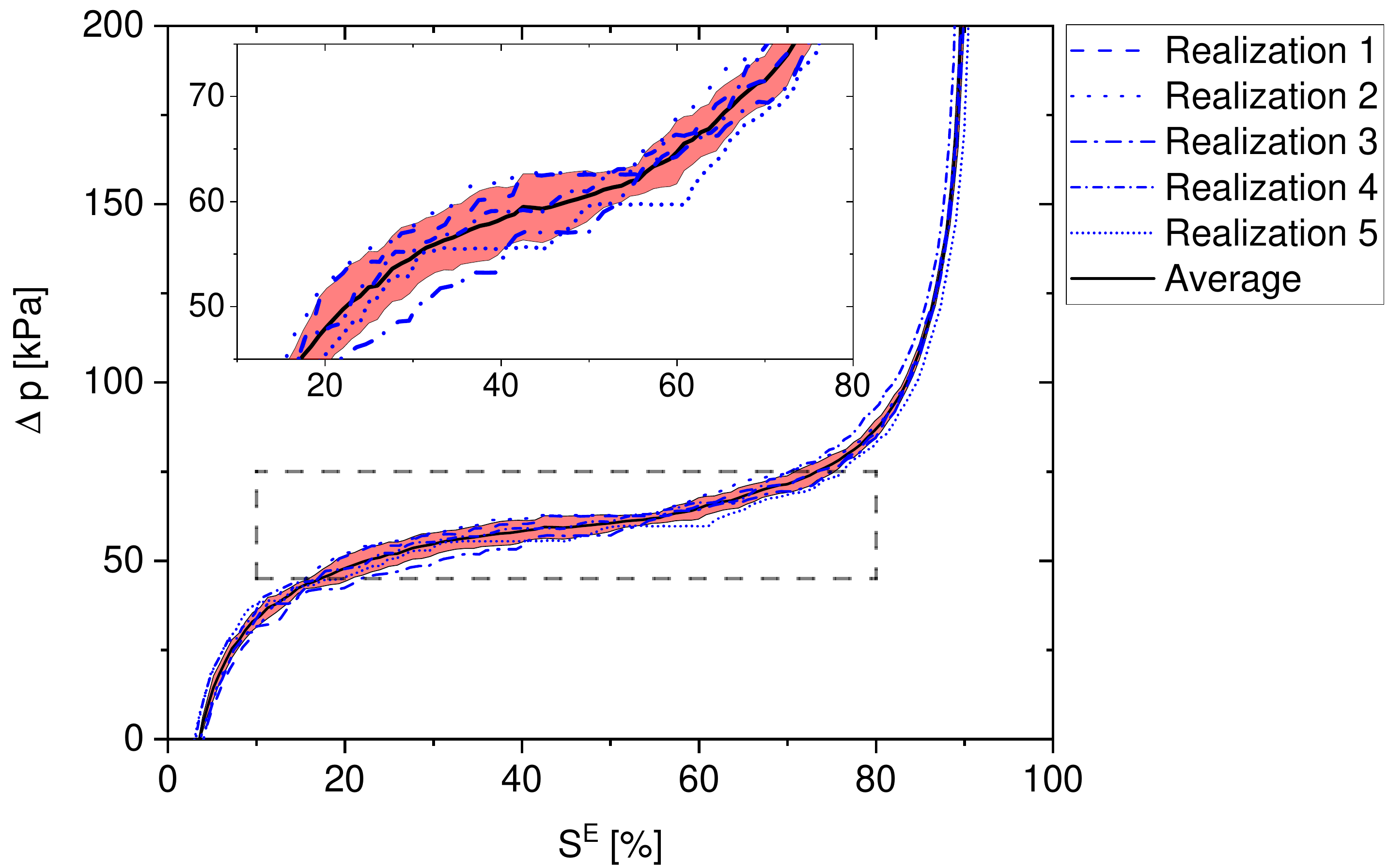}
	\caption{Comparison of pressure-saturation curves determined from filling simulations of five different electrode realizations with similar macroscopic electrode properties ($R_{\mathrm{PS}}=\mathrm{medium}$, $\phi_{\mathrm{A}}=0.6$, $\theta_{\mathrm{A}}=90\degree$, no binder content). The corresponding results are shown by the blue lines. The average pressure-saturation curve is shown by the black solid line. The red shaded area shows the confidence band. For a better overview, the section of the figure which is indicated by the dashed gray frame is enlarged and shown at the top. }
	\label{SI-fig:Uncertainty}
\end{figure}

The corresponding results of both the separate and the average pressure-saturation curves are shown in \autoref{SI-fig:Uncertainty}. They are depicted as blue and black solid lines, respectively. The red shaded area shows the confidence band that arises from adding and subtracting the point-wise standard deviation from the average pressure-saturation curve. The results of the different electrode realizations are in good agreement. The mean standard deviation of the average curve is 3.74\,kPa over the full range of values, i.e.\ $S^\mathrm{E} = [4,\,90] \,\%$, including the steep sides, where small shifts in saturation lead to large deviations. For the reduced range, i.e.\ $S^\mathrm{E} = [10,\,80]\,\%$, where the steep sides are excluded, the mean standard deviation is 2.62\,kPa. The average value of the final degree of saturation is $S^\mathrm{E}_{\mathrm{final}}=S^\mathrm{E}(t_{\mathrm{end}})=90.5\,\%$. The corresponding mean standard deviation is 0.75\,\%.

In addition, by relating each discrete value $p(S^\mathrm{E}_i)$ with $i  \in \{1,...,\mathrm{Data}\}$ to the corresponding value of the average pressure-saturation curve, i.e.\ $\left<\Delta p(S^\mathrm{E}_i)\right>$, also the bias
\begin{align}
    \label{SI-eq:bias}
    \mathrm{bias} = \frac{1}{N_{\mathrm{Data}}} \sum \limits_{i = 1}^{N_{\mathrm{Data}}} \left (1 - \frac{\Delta p(S^\mathrm{E}_i)}{\left<\Delta p(S^\mathrm{E}_i)\right>} \right ),
\end{align}
the average absolute deviation (AAD)
\begin{align}
    \label{SI-eq:AAD}
    \mathrm{AAD} = \frac{1}{N_{\mathrm{Data}}} \sum \limits_{i = 1}^{N_{\mathrm{Data}}} \left |1 - \frac{\Delta p(S^\mathrm{E}_i)}{\left<\Delta p(S^\mathrm{E}_i)\right>} \right |,
\end{align}
and the maximum deviation ($\Delta_{\mathrm{max}}$)
\begin{align}
    \label{SI-eq:Dmax}
    \Delta_{\mathrm{max}} = \max_{i=1,...,N_{\mathrm{Data}}} \left (\left |1 - \frac{\Delta p(S^\mathrm{E}_i)}{\left<\Delta p(S^\mathrm{E}_i)\right>} \right | \right )
\end{align}
were determined. The values are given in \autoref{SI-tab:UncertaintyStatistics}.

\begin{table}[h!]
	\centering
	\caption{Determination of the numerical uncertainty of pressure-saturation curves. Results for five different structural realizations with the same macroscopic electrode parameters were compared. The statistical evaluation is given by the AAD, bias, and maximum deviation. Two sets of comparisons are shown. One spanning the full range of data points, i.e.\ $S^\mathrm{E} = [4,\,90]\,\%$, and another spanning a reduced range, i.e.\ $S^\mathrm{E} = [10,\,80]\,\%$. The mean standard deviations for the full and the reduced range are 3.74\,kPa and 2.62\,kPa, respectively.}
	\begin{tabular}{c|lrrrr}
		\toprule \toprule
             & data range & $N_{\mathrm{Data}}$ & bias (\%) & AAD (\%) & $\Delta_{\mathrm{max}}$ (\%) \\
		\midrule
	 	realization 1    & $S^\mathrm{E} = [4,\,90]\,\%$	& 104 & -0.75 & 3.89 & 23.35 \\
                     & $S^\mathrm{E} = [10,\,80]\,\%$ 	&  76 & -1.86 & 3.11 & 12.29 \\ \hline
	 	realization 2    & $S^\mathrm{E} = [4,\,90]\,\%$	& 104 & -3.01 & 5.99 & 52.18 \\
                     & $S^\mathrm{E} = [10,\,80]\,\%$ 	&  76 & -4.37 & 4.64 &  8.68 \\ \hline
	 	realization 3    & $S^\mathrm{E} = [4,\,90]\,\%$	& 105 &  3.82 & 4.55 & 26.69 \\
                     & $S^\mathrm{E} = [10,\,80]\,\%$ 	&  76 &  4.29 & 4.49 & 10.25 \\ \hline
	 	realization 4    & $S^\mathrm{E} = [4,\,90]\,\%$	& 105 & -1.74 & 3.12 & 39.70 \\
                     & $S^\mathrm{E} = [10,\,80]\,\%$ 	&  76 & -1.27 & 1.70 &  7.39 \\ \hline
	 	realization 5    & $S^\mathrm{E} = [4,\,90]\,\%$	& 105 &  1.64 & 6.09 & 87.88 \\
                     & $S^\mathrm{E} = [10,\,80]\,\%$ 	&  76 &  3.21 & 3.74 &  9.34 \\
		\bottomrule \bottomrule
	\end{tabular}%
	\label{SI-tab:UncertaintyStatistics}%
\end{table}%

Regarding \autoref{SI-tab:UncertaintyStatistics} the AADs of all samples show a good agreement with the average pressure-saturation curve. The best agreement was observed for the electrode realization~4. It was therefore used as a reference and is denoted as default simulation or ID~1 in the present paper.

\section{Numerical Simulation Results from the Present Study} \label{SI-sec:NumericalResults}

The numerical results of the pressure-saturation simulations are summarized in \autoref{SI-file:PressureSaturationData}. In addition, the data of the permeabilities $k^{\mathrm{E}}_y$ and $k^{\mathrm{G}}_y$, the tortuosities $\tau_0$ and $\tau_{\text{end}}$, as well as the electrochemically active surface area $A_{\text{A,act}}$ are given in \autoref{SI-file:PermTortArea}. All data are provided as \textit{.xls}-files and are attached to the Supporting Information.

\begin{file}[h!]
  \centering
	\caption{Data set of the pressure-saturation relationships from the present work. Results are sorted by their simulation ID. Beside the simulation time $t$ also the electrolyte saturation $S^\mathrm{E}(t)$ and the pressure $\Delta p(t)$ is given. The data are  shown in the Figures~2, 3, and 4 in the main paper. \\
	see \textit{PressureSaturation.xls}}
	\label{SI-file:PressureSaturationData}%
\end{file}

\begin{file}[h!]
  \centering
	\caption{Data set of the of the electrolyte and gas permeabilities $k^{\mathrm{E}}_y$ and $k^{\mathrm{G}}_y$ as well as their standard deviations. In addition, the data of the geodesic tortuosities $\tau_0$ and $\tau_{\text{end}}$ as well as the electrochemically active surface area $A_{\text{A,act}}$ are given. The data are shown in Figures~8, 9, and 10 in the main paper. \\
	see \textit{Permeability\_Tortuosity\_ActiveArea.xls}}
	\label{SI-file:PermTortArea}%
\end{file}

\section{Total Duration of the Filling Process} \label{SI-sec:Duration} 

\begin{figure}
	\centering
	  \includegraphics[width=0.6\textwidth]{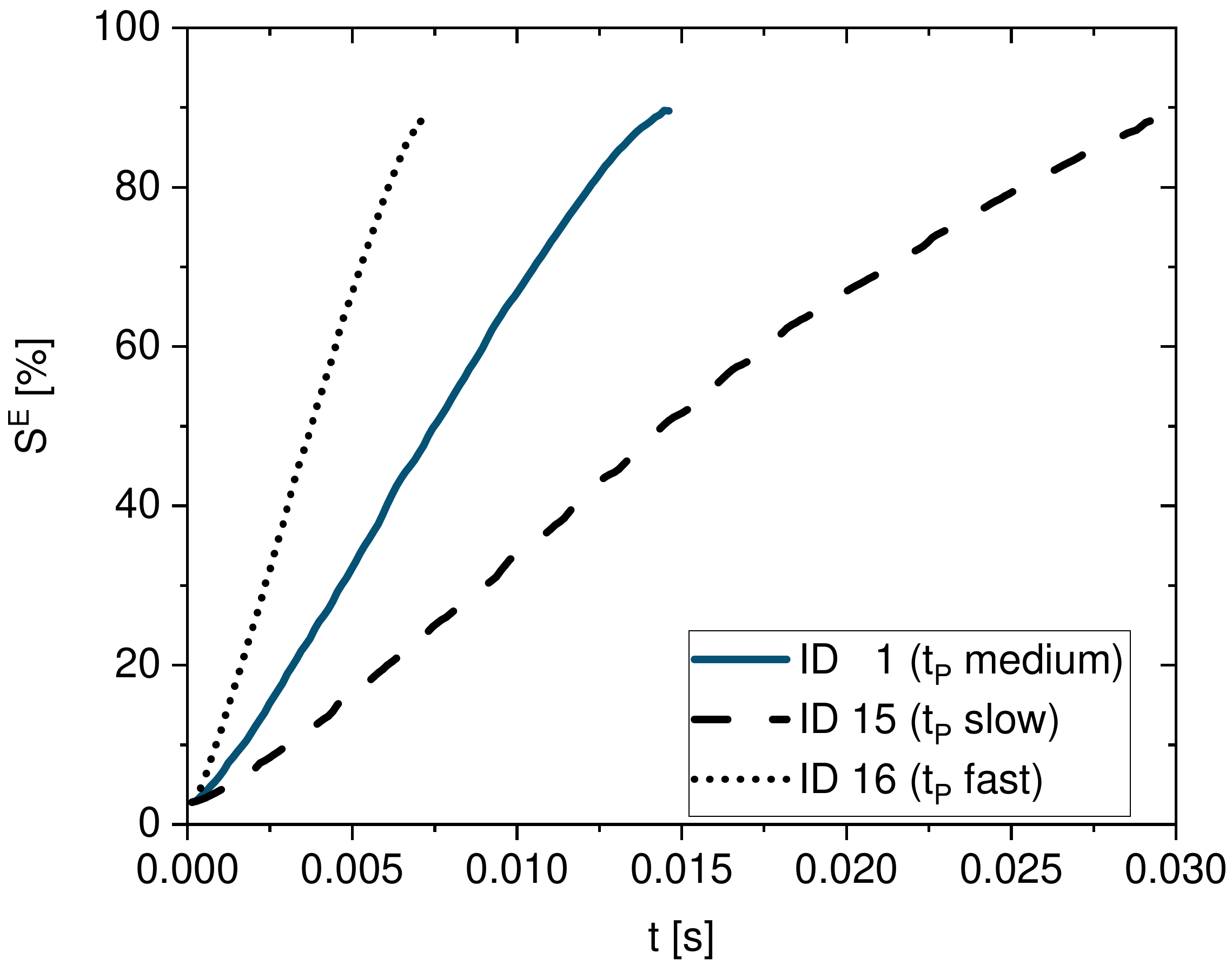}
	\caption{Saturation-time behavior for the reference electrode without binder. Results are shown for three processing times: ID~1 ($t_{\text{P}}=\text{medium}$, blue solid line), ID~15 ($t_{\text{P}}=\text{slow}$, black dashed line), and ID~16 ($t_{\text{P}}=\text{fast}$, black dotted line).}
	\label{SI-fig:SaturationTime}
\end{figure}

In Section 5.2 of the present paper the process times of the filling processes are discussed. The corresponding saturation-time behaviors are shown in \autoref{SI-fig:SaturationTime} in the follwing. 

\section{Gas Entrapment} \label{SI-sec:GasEntrapment} 

\begin{figure}
	\centering
	  \includegraphics[width=1\textwidth]{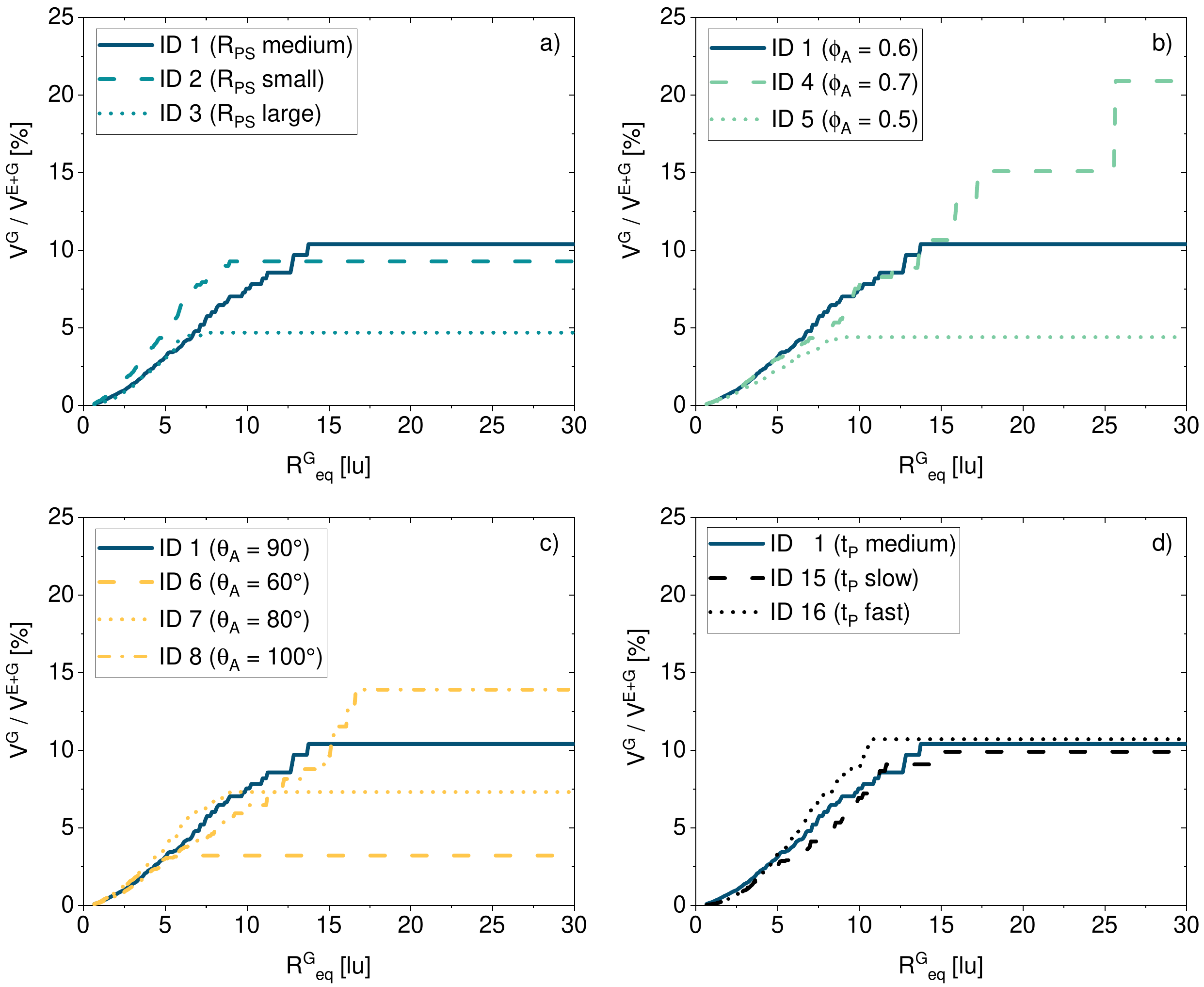}
	\caption{Relation between the cumulated gas volume $V^{\mathrm{G}}$ divided by the total pore volume $V^{\mathrm{E+G}}$ and the equivalent gas bubble radius $R^{\mathrm{G}}_{\mathrm{eq}}$. Results are shown for the IDs~1$-$8 and 15$-$16. The default simulation, i.e.\ ID~1, is depicted with the blue solid line. The influencing factors are indicated by the colors. Those are a) the particle size distribution $R_{\mathrm{PS}}$ (turquoise), b) the volume fraction of the active material $\phi_{\mathrm{A}}$ (green), c) the wettability $\theta_{\mathrm{A}}$ (orange), and d) the process time $t_{\mathrm{P}}$ (black).}
	\label{SI-fig:cumulatedVolume_noBinder}
\end{figure}

In Section 5.3 of the present paper the gas entrapment at the end of the filling process is discussed. However, the corresponding size distributions of the gas agglomerates are only exemplarily shown for the reference cases ID~1 and ID~9. In \figsref{SI-fig:cumulatedVolume_noBinder}{SI-fig:cumulatedVolume_Binder} they are shown for all simulations, i.e.\ ID~1$-$16. The results are given as the ratio of the cumulated gas volume $V^{\mathrm{G}}$ to the total pore volume $V^{\mathrm{E+G}}$. They are plotted as a function of the equivalent gas bubble radius $R^{\mathrm{G}}_{\mathrm{eq}}$. 

\begin{figure}
	\centering
	  \includegraphics[width=1\textwidth]{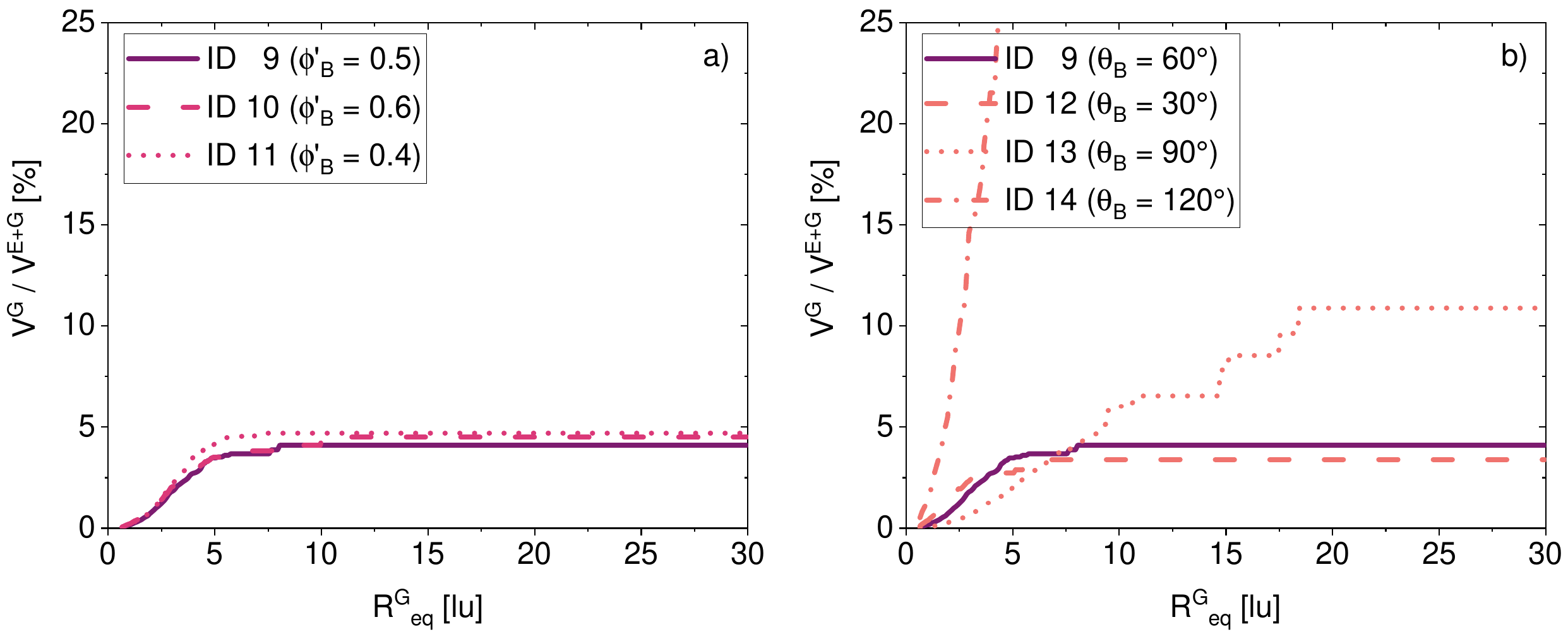}
	\caption{Relation between the cumulated gas volume $V^{\mathrm{G}}$ divided by the total pore volume $V^{\mathrm{E+G}}$ and the equivalent gas bubble radius $R^{\mathrm{G}}_{\mathrm{eq}}$. Results are shown for the IDs~9$-$14. The default simulation with binder, i.e.\ ID~9, is depicted with the purple solid line. The influencing factors are indicated by the colors. Those are a) the inner volume fraction of the binder $\phi'_{\mathrm{B}}$ (magenta), and b) the wettability $\theta_{\mathrm{B}}$ (red). }
	\label{SI-fig:cumulatedVolume_Binder}
\end{figure}

\end{document}